%% file: R2_arXiv_MAIN.tex
\documentclass[fleqn,10pt]{Setup_files/wlscirep}

\usepackage{float}
\usepackage{framed}

\usepackage{bm}
\usepackage{physics}
\usepackage{amsfonts}
\usepackage[utf8]{inputenc}
\usepackage[T1]{fontenc}
\usepackage{mathpazo} 
\usepackage{amsmath}
\usepackage{textalpha}
\usepackage{lineno}
\usepackage[normalem]{ulem}
\input{Setup_files/Definitions}

\newcommand{\revv}[1]{\textcolor{black}{#1}}
\newcommand{\rev}[1]{\textcolor{black}{#1}}

\title{Task-adaptive physical reservoir computing}

\author[1,*]{Oscar Lee}
\author[1]{Tianyi Wei}
\author[2]{Kilian D. Stenning}
\author[2]{Jack C. Gartside}
\author[1]{\rev{Dan Prestwood}}
\author[3]{Shinichiro Seki}
\author[4,5]{Aisha Aqeel}
\author[6]{\rev{Kosuke Karube}}
\author[3]{\rev{Naoya Kanazawa}}
\author[6]{\rev{Yasujiro Taguchi}}
\author[4]{Christian Back}
\author[3,6,7]{Yoshinori Tokura}
\author[2,8]{Will R. Branford}
\author[1,9,10,**]{Hidekazu Kurebayashi}

\affil[1]{London Centre for Nanotechnology, University College London, London, WC1H 0AH, United Kingdom}

\affil[2]{Blackett Laboratory, Imperial College London, London, SW7 2AZ, United Kingdom}

\affil[3]{Department of Applied Physics, University of Tokyo, Tokyo, 113-8656, Japan}

\affil[4]{Physik-Department, Technische Universitat Munchen, Garching, D-85748, Germany}

\affil[5]{Munich Center for Quantum Science and Technology (MCQST), Munich, D-80799, Germany}

\affil[6]{\rev{RIKEN Center for Emergent Matter Science (CEMS), Wako, 351-0198, Japan}}

\affil[7]{Tokyo College, University of Tokyo, Tokyo, 113-8656, Japan}

\affil[8]{London Centre for Nanotechnology, Imperial College London, London, SW7 2AZ, United Kingdom}

\affil[9]{Department of Electronic and Electrical Engineering, University College London, London, WC1E 7JE, United Kingdom}

\affil[10]{WPI Advanced Institute for Materials Research, Tohoku University, 2-1-1, Katahira, Sendai 980-8577, Japan}

\affil[*]{e-mail: s.lee.14@ucl.ac.uk}
\affil[**]{e-mail: h.kurebayashi@ucl.ac.uk}

\begin{abstract}
Reservoir computing is a neuromorphic architecture that potentially offers viable solutions to the growing energy costs of machine learning.
\revv{In software-based machine learning,} neural network properties and performance can be \revv{readily} reconfigured to suit different computational tasks \revv{by changing hyperparameters}.
This critical functionality is missing in ``physical" reservoir computing schemes that exploit nonlinear and history-dependent memory responses of physical systems for data processing. Here, we experimentally present a `task-adaptive' approach to physical reservoir computing, capable of reconfiguring key reservoir properties (nonlinearity, memory-capacity and complexity) to optimise computational performance across a broad range of tasks. 
As a model case of this, we use the temperature and magnetic-field controlled spin-wave response of Cu$_2$OSeO$_3$ that hosts skyrmion, conical and helical magnetic phases, providing on-demand access to a host of different physical reservoir responses. We quantify phase-tunable reservoir performance, characterise their properties and discuss the correlation between these in physical reservoirs. 
This task-adaptive approach overcomes key prior limitations of physical reservoirs, opening opportunities to apply thermodynamically stable and metastable phase control across a wide variety of physical reservoir systems\rev{, as we show its transferable nature using above(near)-room-temperature demonstration with Co$_{8.5}$Zn$_{8.5}$Mn$_{3}$ (FeGe)}.
\end{abstract}

\begin{document}

\flushbottom
\maketitle

\section*{Introduction}

Physical separation between processing and memory units in the current computer architecture causes significant energy waste due to repeated shuttling of data, known as the von Neumann bottleneck. To circumvent this, neuromorphic computing~\cite{roy_NaturePerspective2019,markovic2020physics,schuman_NatureComputationalScience2022}, which emulates the brain's neural network to co-locate memory and processor to integrated `memcomputing' units, has attracted a great deal of attention as a promising future technology for artificial intelligence processing. 
Reservoir computing~\cite{jaeger2001echo,Maass_NeuralComm2002,Steil_IEEE2004,nakajima2021reservoir} is a type of neuromorphic architecture with complex recurrent pathways (the `reservoir') that map input data to a high-dimensional space. Weights within the reservoir are randomly initialised and fixed, and only the small one-dimensional weight vector that linearly connects the reservoir to the output requires optimisation using a computationally-cheap linear regression. 
As such, reservoir computing can achieve powerful neuromorphic computation at a fraction of the processing cost relative to other schemes, e.g. deep neural network, where the whole (typically more than millions of) weight network must be trained~\cite{Jaeger_GMDRepoert2002}.

While reservoir computing was originally conceived in software~\cite{jaeger2001echo}, nonlinear and history-dependent responses of physical systems have also been exploited as reservoirs~\cite{Kulkarni_IEEE2012,OBST_NanoCommNet2013,tanaka2019recent}. The field of physical reservoir computing has been rapidly expanding with several promising demonstrations using optical systems~\cite{Duport_OptExp12}, analogue electronic circuits~\cite{Soriano_IEEE2014}, memristors~\cite{Du_NatComm2017,Moon_NatureElectron2019}, ferroelectrics~\cite{Liu_AdvMater2022}, magnetic systems~\cite{grollier2020neuromorphic,nakane2018reservoir,Nomura_JJAP2019,Tsunegi_APL2019,Gartside_naturenano2022,Allwood_APL_2023,Dawidek_Wiley_2021} and even a bucket of water~\cite{Fernando_Springer2003}. 
Skyrmions, topologically non-trivial magnetic whirls, have also been proposed as hosts for reservoir computing~\cite{Prychynenko_PRAppl2018,Pinna_PRAppl2020,msiska2022audio,lee2023perspective} as part of rapidly growing research efforts towards neuromorphic computing~\cite{torrejon2017neuromorphic,Zazvorka_NatNano2019,Song_NatElec2020,Zahedinejad_NatMater2022,papp2021nanoscale}. Very recently, there have been a few experimental studies using skyrmions in metallic multilayers to perform reservoir computing by using electric read-out (anomalous Hall resistance)~\cite{Yokouchi_SciAdv2022} or skyrmion position measured by Kerr microscopy~\cite{Raab_arXiv2022}.

Despite the rapid development, one of the major outstanding challenges for creating powerful physical reservoirs is establishing a methodology for task-adaptive control of reservoir properties~\cite{tanaka2019recent}, often characterised by the nonlinearity, memory-capacity and complexity metrics of the reservoir~\cite{lukovsevivcius2009reservoir,dambre_scientificReports2012,Inubushi_ScienctificReports2017,Love_arXiv2021,stenning2023adaptive}. However, physical systems typically have a narrow and fixed set of reservoir properties without having much room to change, as the above metrics tend to be constrained to a particular response phenomenon of a physical system. This creates challenges where a physical reservoir may perform well for some specific tasks, but poorly at others which require different reservoir properties~\cite{John_NatComms2022}. This is a severe drawback relative to software reservoirs, where such properties can be tuned by changing lines of code~\cite{Joy_arXiv2022}.

\begin{figure}
    \centering
    \includegraphics[width=1\linewidth]{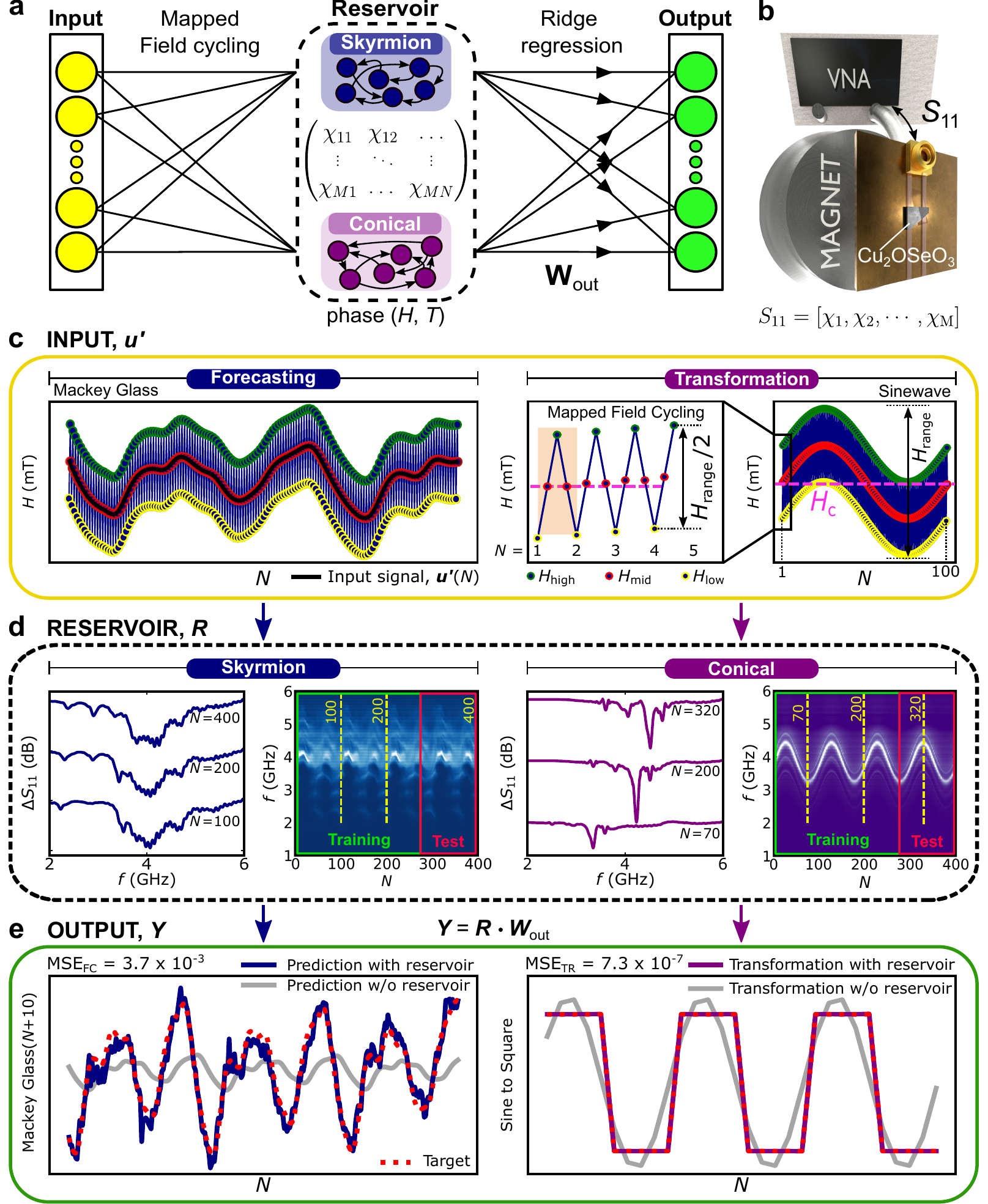}
\end{figure}
\addtocounter{figure}{0}
\begin{figure}
  \caption{\textbf{a}, Illustration of a task-adaptive reservoir computing framework. Different magnetic phases are accessed by controlling the external field ($H$) and temperature ($T$). \textbf{b}, Experimental schematic of VNA-assisted spin-wave spectroscopy setup. \textbf{c}, Typical input scheme for forecasting (left: Mackey-Glass signal) and transformation (right: sinewave) tasks. The original input signal, \textbf{\textit{u}}(\textit{t}), is mapped to \textbf{\textit{u’}}(\textit{N}), defined by the mapped field-cycling protocol (details in main text). Note that \rev{$H_{\rm range}$ defines the range of applied fields, where,} the distance between $H_{\rm low}$ and $H_{\rm high}$ is the width of cycling, $H_{\rm range}$\rev{/2}. A single field-cycle is highlighted by the orange box in the right panel. \textbf{d}, $S_{11}$ as a function of $f$ after accumulating \textit{N} field-cycles and visualisation of $\textbf{\textit{R}}$; a collective spectral evolution for \textit{N} field-cycles for skyrmion and conical phases, separated into ``training” and ``test” datasets. \textbf{e}, Results after applying $\textbf{W}_{\rm out}$ on the unseen ``test” dataset. Left: Forecasting of a differential chaotic time series data, Mackey-Glass signal by 10 future steps. Right: Transformation of a sinewave to a square wave signal. In both cases, reservoir prediction (transformation) results are plotted in blue (purple), the red-dotted line depicts the target signal, and the grey line represents the control prediction where ridge regression is performed on the raw input data without the physical reservoir. MSE$_{\rm FC}$ and MSE$_{\rm TR}$ quantify the computation performance of forecasting and transformation, respectively.}
  \label{Fig:1}
\end{figure}

Here, we demonstrate task-adaptive physical reservoir computing using the spectral space of a physical system that has rich, phase-tunable dynamical modes. As a model system of this approach, we use spin resonances of the chiral magnet Cu$_2$OSeO$_3$~\cite{seki_science2012,Onose_PRL2012,Garst_JPhysD2017}. Since different magnetic phases (skyrmion, helical and conical) exhibit distinct resonant dynamics, the phases offer broadly varying reservoir properties and computing performance, which can be reconfigurably tuned via magnetic field and temperature. We use magnetic field-cycling~\cite{Aqeel_PRL2021,Lee_JPhysCondMat2021} to input data and measure spin-wave spectra at each input step to efficiently achieve high-dimensional mapping by exploiting the spectral response of each magnetic mode. By quantitatively assessing each reservoir phase, we find that the \revv{thermodynamically metastable} skyrmion phase has a strong memory-capacity due to \revv{magnetic-field-driven} gradual \revv{nucleation} of \revv{skyrmions} with excellent performance in future prediction tasks. In contrast, the conical phase has modes with great reservoir nonlinearity and complexity, ideal for transformation tasks. By making full use of this phase-tunable nature within a single physical system, we achieve strong performance across a broad range of tasks in a single physical system. Furthermore, we perform a correlation analysis between the reservoir performance quantified by mean squared error (MSE) and the reservoir properties; nonlinearity, memory-capacity and complexity. \rev{High temperature demonstration of the task-adaptive physical reservoir concept using other chiral magnets, Co$_{8.5}$Zn$_{8.5}$Mn$_{3}$ and FeGe, indicates that the concept is indeed ubiquitous.}

\section*{Working principle of chiral magnet Cu$_2$OSeO$_3$ physical reservoir}

Our physical reservoir (Fig.~\ref{Fig:1}a) is constructed using field- and temperature-dependent GHz spin dynamics of Cu$_2$OSeO$_3$~\cite{Onose_PRL2012}. Similar to a recent reservoir computing methodology reported by several co-authors~\cite{Gartside_naturenano2022}, we apply a specific sequence of magnetic field inputs and map out the spin-wave spectra of Cu$_2$OSeO$_3$ to form a two-dimensional matrix. Subsequently, the reservoir matrix is multiplied by a weight vector $\textbf{W}_{\rm out}$ to produce the individual output value for each input. We use standard ridge regression to train/calculate $\textbf{W}_{\rm out}$ for each task with training data. The trained reservoir is then run for the unseen data (test) sets to assess the reservoir computing performance via MSE (see supplementary materials (SM) Section 2 for more details). The rich phase diagram of Cu$_2$OSeO$_3$ offers multiple magnetic textural phases, including the thermodynamically metastable skyrmion phase~\cite{oike_naturephysics2015,Qian_ScienceAdvances2018,Halder_PRB2018,Chacon_naturephysics2018,Aqeel_PRL2021,Lee_JPhysCondMat2021}, each exhibiting distinct spin-dynamics properties. 

The task-adaptive nature of our physical reservoir comes from the reconfigurable on-demand control over balancing between these stable and metastable magnetic phases by both temperature and magnetic field. For our experiments, a polished plate-shaped bulk Cu$_2$OSeO$_3$ crystal of dimensions 1.9, 1.4 and 0.3~mm (x, y, z) was placed on a coplanar waveguide with (100) surface facing down, where the microwave reflection spectra, $S_{11}$, were recorded as a function of frequency $f$ and field using a vector network analyser (VNA) as depicted in Fig.~\ref{Fig:1}b. 

Next, we describe our reservoir computing process which has three components: input, reservoir and output. The input layer consists of sequential field values, \textbf{\textit{u’}}~=~($H_1$, $H_2$, $H_3$, ..., $H_n$), \revv{produced by projecting an input function into a magnetic field value as summarised in Fig.~\ref{Fig:1}c.} \revv{Taking the transformation task as an example, each} field-cycle $N$ starts with a low magnetic field $H_{\rm low}$, increasing to a high magnetic field $H_{\rm high}$ and comes back to a new $H_{\rm low}$, where their separation is defined by $H_{\rm range}$\rev{/2} \revv{with a centre field $H_{\rm c}$}. The individual field points ($H_{\rm low}$, $H_{\rm mid}$ and $H_{\rm high}$) are modulated by the input functions tailored for specific tasks. For example, for transformation tasks, the input function is a sine curve encoded over 100 field-cycles; our forecasting tasks use \revv{a chaotic oscillatory Mackey-Glass time series~\cite{Mackey_1977}} to modulate the field-cycling base with $N$ \revv{as shown in the left panel in Fig.~\ref{Fig:1}c} (see more details in SM Section 2). This scheme can be applied to input any time series dataset into the physical reservoir. To create a two-dimensional reservoir matrix, $S_{11}$($N$,~$f$), we measure the reflection coefficient spectra $S_{11}$ consisting of $M$ frequency-channels (here 1601) between 1 and 6~GHz for each field-cycle at $H_{\rm low}$ labelled by $N$ (see SM \rev{Section 1} for more details). As such, the physical reservoir effectively broadcasts a \revv{single field input value} to 1601 outputs via ferromagnetic resonance frequency-multiplexing.

Figure~\ref{Fig:1}d shows the spectral output of our reservoir in response to input time series datasets (left: Mackey-Glass, right: sinewave). The spectral states of each phase (left: skyrmion, right: conical) change as we perform field-cycling - see individual spectra sampled at different $N$ values in Fig.~\ref{Fig:1}d. By using $S_{11}(N, f)$ in the colour heatmap plots, we form the reservoir matrix, \textbf{\textit{R}}, comprising $M$ rows and $N$ columns as shown in the middle panel of Fig.~\ref{Fig:1}a where $\chi_{\rm ij}$ represents the magnetic susceptibility for each input field and frequency. 

Using 70~\% of the reservoir response as the train dataset $\textbf{\textit{R}}_{\rm train}$ shown in Fig.~\ref{Fig:1}d, we perform ridge regression to calculate the weights $\textbf{W}_{\rm out}$ against a target function $\textbf{\textit{Y}}$:~$\textbf{\textit{Y}}$~=~$\textbf{\textit{R}}_{\rm train}$~$\cdot$~$\textbf{W}_{\rm out}$ which represents the desired task. The calculated $\textbf{W}_{\rm out}$ and the remaining 30~\% of the reservoir $\textbf{\textit{R}}_{\rm test}$ are subsequently used to evaluate reservoir performance quantitatively via MSE. Figure~\ref{Fig:1}e exemplifies this final process of our reservoir computing protocol by showing the physical reservoir's attempt (blue line) at reproducing the target signal (red dotted line) for two tasks: left, a forecast of the chaotic Mackey-Glass signal 10 future steps ahead and right, a nonlinear transformation of a sinewave input to a square wave target. For both tasks, excellent performances of reservoir computing are confirmed by low MSE values; 3.7$\times$10$^{-3}$ for the forecasting task by the skyrmion reservoir and 7.3$\times$ 10$^{-7}$ for the transforming task by the conical reservoir. The significance of reservoir components can be assessed by these two values with those calculated by computing the same tasks without the reservoirs, 6.2$\times$10$^{2}$ and 5.4$\times$10$^{2}$ for the forecasting and transformation tasks respectively. 

\section*{Phase-tunable physical reservoir computing}

\begin{figure}[h!]
\centering
\includegraphics[width=1\linewidth]{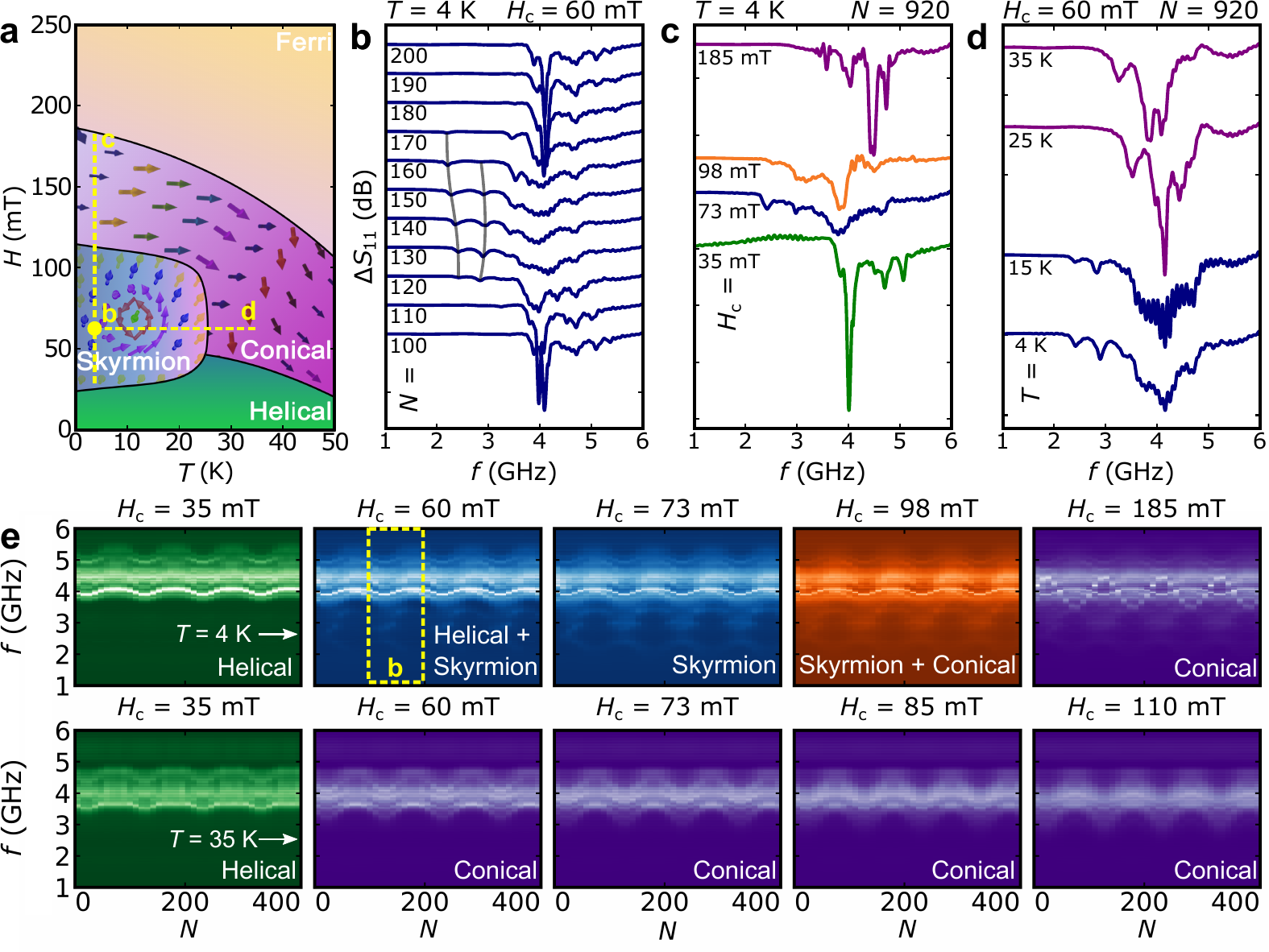}
\caption{\textbf{a}, Schematic of the temperature-phase diagram for a bulk crystal Cu$_2$OSeO$_3$. The yellow dotted vertical (horizontal) line indicates the experimental conditions for our cycling experiments shown in \textbf{c}(\textbf{d}). \textbf{b}, The cycling number dependence of spin-wave spectra in Cu$_2$OSeO$_3$ for $H_{\rm c}$~=~60~mT and 4~K. The evolution of the skyrmion-phase spectra is shown for increasing values of \textit{N}. Grey lines are added in (\textbf{b}) as a guide to the eye to keep track of the skyrmion modes. \textbf{c}, $H_{\rm c}$ dependence of spin-wave spectra in Cu$_2$OSeO$_3$ for 4~K and after 920 field-cycles. \textbf{d}, Temperature dependence of the spin-wave spectra for $H_{\rm c}$~=~60~mT after 920 field-cycles. \textbf{e}, Microwave absorption spectra as a function of $f$ and $N$ for different values of $H_{\rm c}$ at $T=$~4~K (upper row) and 35~K (lower row). The input signal in all plots is a sinewave with $H_{\rm range}$~=~90~mT.}
\label{Fig:2}
\end{figure}

The phase-tunable nature of our physical reservoir computing stems from Cu$_2$OSeO$_3$'s rich magnetic phase-diagram shown in Fig.~\ref{Fig:2}a~\cite{seki_science2012}. Added to this diagram is the metastable skyrmion phase, which can be generated at low temperatures below $\sim$35~K by quenching techniques or field-cycling protocols~\cite{oike_naturephysics2015,Lee_JPhysCondMat2021,Aqeel_PRL2021}. We leverage this phase-tunability to create the task-adaptive nature of our physical reservoir as detailed below. 

Figure~\ref{Fig:2}b displays the cycle-number dependence of the spectra for $H_{\rm c}$ and temperature inside the skyrmion phase. For $N$~=~100, a sharp peak around 4~GHz can be clearly observed, corresponding to \revv{low-energy spin-wave} modes of the thermodynamically stable conical phase~\cite{Garst_JPhysD2017,Lee_JPhysCondMat2021,Aqeel_PRL2021}. As we cycle further, the \revv{conical} mode amplitude is shrunk, and the skyrmion modes appear around 2~-~3~GHz as highlighted by grey curves for $N$~=~130-170. These are the counter-clockwise and breathing modes of the metastable low-temperature skyrmion phase generated by field-cycling~\cite{Lee_JPhysCondMat2021,Aqeel_PRL2021}. The mode \revv{frequencies move with our input magnetic fields and as} the cycling \revv{proceeds}, \revv{skyrmions are} continuously destroyed and renucleated\revv{, evident by the peak amplitude}. When we carry out \revv{experiments} for different $H_{\rm c}$, we can clearly demonstrate the tunability of magnetic phases for our reservoir computing as we show in Fig.~\ref{Fig:2}c, where the spectra are taken after 920 field-cycles with $H_{\rm range}$~=~90~mT at 4~K. A similar tunability can be achieved by changing temperature at a fixed $H_{\rm c}$ of 60~mT as shown in Fig.~\ref{Fig:2}d. The skyrmion modes are clearly identified for 4~K and 15~K and disappear for higher temperatures 25~K and 35~K, where the spectra are dominated by multiple broad modes between 3~-~5~GHz from the conical phase. Finally, a collection of the field-cycle evolution of spectra for various $H_{\rm c}$ and temperatures are shown in Fig.~\ref{Fig:2}e to demonstrate the range of phase/spectral tunability\rev{. Individual spectral scans for \rev{further evolution of $N$  as a variation of \Hc{}} can be found in SM Section 3}.

\section*{Reservoir Performance}

Figures~\ref{Fig:3}a-c compare the reservoir's performance on different tasks using magnetic phases of skyrmion ($H_{\rm c}$~=~60~mT), skyrmion-conical hybrid ($H_{\rm c}$~=~98~mT) and conical modes ($H_{\rm c}$~=~185~mT) at 4~K with $H_{\rm range}$~=~90~mT and $N$~=~1000. For forecasting, the system is trained to predict the future behaviour of a Mackey-Glass signal of 10 steps ahead. Reservoir performance is evaluated quantitatively by calculating MSE between the reservoir prediction and the target signal.

\begin{figure*}
\centering
\includegraphics[width=1\linewidth]{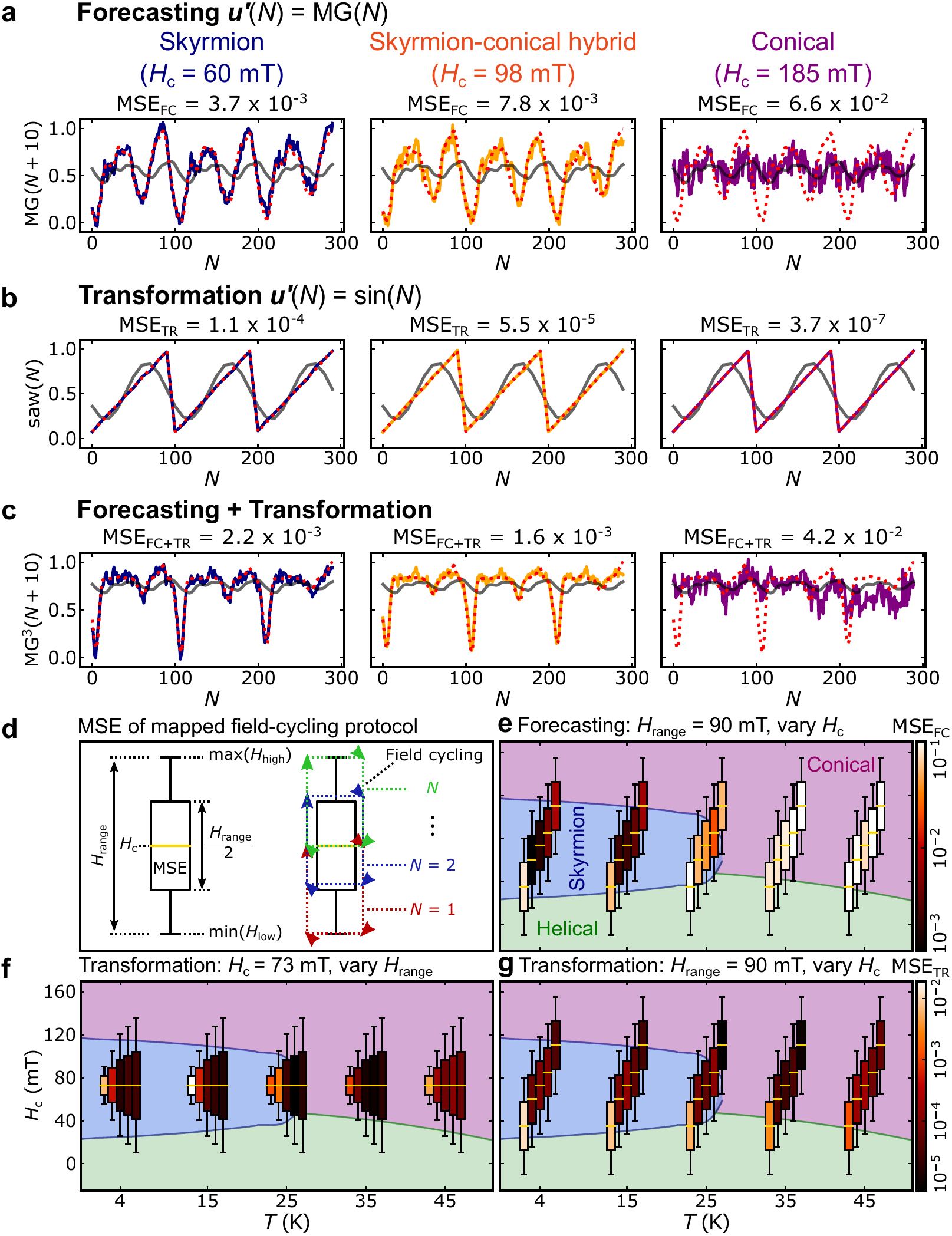}
\end{figure*}
\addtocounter{figure}{0}
\begin{figure}
  \caption{MSE performance comparison of different computation tasks across three distinct physical phases (skyrmion, hybrid and conical) at $T=$~4~K. In these figures, red dotted and grey curves represent the target functions and computation results without the physical reservoirs. Blue, orange, and purple curves display calculations with the physical reservoirs of skyrmion, hybrid and conical phases, respectively. \textbf{a}, Forecasting a Mackey-Glass chaotic time series by 10 future steps (MG($N+10$)). \textbf{b}, Nonlinear transformation of a sinewave input into saw waveforms. \textbf{c}, Combined transform/forecasting of 10 future steps of a cubed Mackey-Glass signal. \textbf{d}, Illustration of the mapped field-cycling protocol visualised as a boxplot (details in main text). \textbf{e\&g}, Evaluation of MSE values at a constant $H_{\rm range}$ as a variation of $H_{\rm c}$ and \textit{T}, respectively, for forecasting (MG($N+10$)) and transformation (square wave) target applications. \textbf{f}, Evaluation of MSE values at a constant $H_{\rm c}$ as a variation of $H_{\rm range}$ and \textit{T} for a transformation (square wave) target application.}
  \label{Fig:3}
\end{figure}

As shown in Fig.~\ref{Fig:3}a, when $H_{\rm c}$ increases and the reservoir is transfigured from the skyrmion to conical phase, prediction performance deteriorates and MSE increases by approximately a factor of 18. In the conical phase, the reservoir prediction is as bad as the one without the reservoir. The opposite trend is observed for transformation tasks, where MSE is significantly improved when switching from the skyrmion reservoir to the conical reservoir as shown in Fig.~\ref{Fig:3}b. While the skyrmion reservoir still performs well with MSE in orders of $10^{-4}$, the conical reservoir excels with the MSE of $3.7~\times 10^{-7}$ for the sine-to-saw transformation task. By setting $H_{\rm c}$ at 98~mT, we create a hybrid reservoir phase where both skyrmion and conical modes coexist. This particular reservoir configuration outperforms both individual skyrmion or conical reservoirs for a complex task combining forecasting and transformation, predicting 10 future steps ahead (forecasting) for a cubed (transformation) Mackey-Glass signal from a normal Mackey-Glass input shown in Fig.~\ref{Fig:3}c. \rev{See SM Section 5 for details of target generations, and a broader selection of further} forecasting and transformation tasks with strong reservoir performance demonstrated throughout.

To map the observed reservoir performance trends across a wider parameter space of $H_{\rm c}$, $H_{\rm range}$ and temperatures, systematic reservoir computing experiments for different reservoir properties across the temperature-field phase diagram were performed as shown in Figs.~\ref{Fig:3}e-g. Figure~\ref{Fig:3}d defines field-cycling parameters to aid reading Figs.~\ref{Fig:3}e-g. The upper and lower whiskers represent the maximum and minimum magnetic field values in the cycling scheme, respectively. The height of the box represents $H_{\rm range}$, and the central line defines $H_{\rm c}$. The MSE values are encoded as the box colour. The initial cycle begins at the bottom of the lower whisker and gradually moves up and down as a function of $N$. Figure~\ref{Fig:3}e shows reservoir performance for forecasting Mackey-Glass($N$~+~10) at $H_{\rm range}$~=~90~mT as a variation of $H_{\rm c}$ and temperatures. The best forecasting performance is found when the field-cycling lies entirely inside the skyrmion phase at lower temperatures. The performance monotonically worsens as field-cycling moves beyond the skyrmion phase and dramatically reduces when leaving the skyrmion phase at high temperatures. The excellent performance of the skyrmion reservoir is highly correlated with its memory-capacity as we discuss \revv{below}. 

For the transformation tasks, we show reservoir performance for two parameter dependencies, $H_{\rm range}$ and $H_{\rm c}$. In Fig.~\ref{Fig:3}f, where a variation of $H_{\rm range}$ for $H_{\rm c}$~=~73~mT is shown, it is clear for all measured temperatures that larger $H_{\rm range}$ values provide optimal reservoir performance, maximising the balance between the key reservoir properties associated with the tasks. In Fig.~\ref{Fig:3}g, we observe that reservoirs run with input mappings extending deeper into the helical phase ($H_{\rm c}$~=~35~mT) perform significantly worse for each temperature measured. In this condition, the field-cycling range crosses the zero-field boundary where the nucleation of the skyrmion modes is reset, suppressing their contributions to the reservoir performance. Optimal performance for the transformation task is demonstrated when the reservoir substantially includes the conical phase that has strong nonlinearity and complexity. The MSE values displayed in Figs.~\ref{Fig:3}e\&g (where the input variation of $H_{\rm c}$ is the same) highlight that performance from the identical reservoirs is starkly different between two types of computational tasks.

\begin{figure}
\centering
\includegraphics[width=1\linewidth]{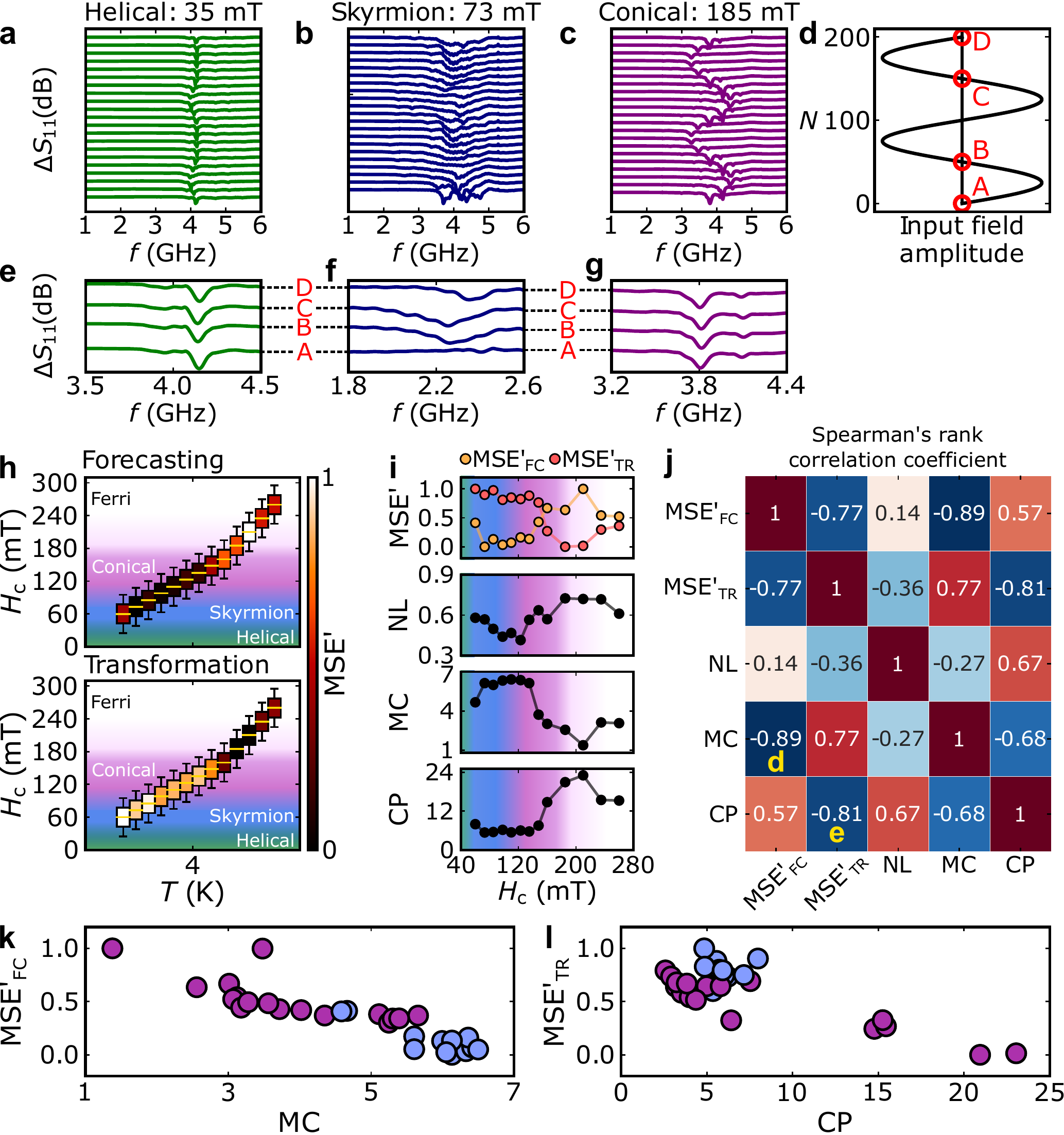}
\caption{\textbf{a-c}, Spin-wave spectra of helical, skyrmion, and conical magnetic phases, left to right respectively. \textbf{d}, Sinewave input sequence defining applied field amplitudes. \textbf{e-g}, Spin-wave spectra at nodes of the sinewave input-fields from \textbf{d}. \textbf{h}, $H_{\rm c}$ evolution of MSE$'$s at $T=$~4~K and $H_{\rm range}$~=~90~mT, for forecasting (MG($N~+~10$)) and transformation (square wave) target applications, respectively. Note that MSE$'$ denotes the normalised scale of MSE between [0, 1], where 0(1) represents the best(worst) MSE. A (meta)stable magnetic field range for each phase is colour-coded. \textbf{i}, MSE$'$ and task agnostic metric results as a function of $H_{\rm c}$ at $T=$~4~K. \textbf{j}, Correlation matrix of Spearman’s rank correlation coefficient. \textbf{k}, Performance of forecasting as an evolution of MC. \textbf{l}, Performance of transformation as an evolution of CP. Here, the colour of the dots represent the corresponding magnetic phase at which the metric was evaluated (blue: skyrmion, purple: conical).}
\label{Fig:4}
\end{figure}

\revv{The computational performance of our magnetic reservoirs can be related to their physical properties. Figures~\ref{Fig:4}a-c display the spectral evolution of different magnetic phases with field-cycling. High(low) transformation performance of the conical(helical) phase can be associated with the size of frequency shift by magnetic field. The dispersion curve of the helical phase displays a notably flat profile in comparison to other magnetic phases in chiral magnets~\cite{Schwarze_NatMater2015}, resulting in poor computational performance with its peak position shifting very weakly in response to field input. Much higher amplitude frequency shifts are found in the highly-performing conical and skyrmion phases, producing the strong nonlinearity and complexity in their reservoirs, hence low MSEs in transformation tasks - see further/detailed analysis in SM Section 4. The origin of excellent performance of the skyrmion reservoirs for forecasting tasks can be explained by comparing the spectra across the three phases at the same field values but different points in the input field cycle, labelled as A-D in Fig.~\ref{Fig:4}d. The spectra of both helical (Fig. 4e) and conical (Fig. 4g) phases are identical across Points A-D, showing that these phases respond only to the current field-input being applied and lack any memory response for magnetic field inputs. In contrast, the skyrmion spectra in Fig.~\ref{Fig:4}f are dissimilar across Points A-D, meaning that the spectral response depends on not only the field value but also past field inputs. This is the source of the crucial physical memory response for forecasting tasks, arising from magnetic field-driven nucleation of metastable skyrmions and annihilation of other magnetic phases~\cite{oike_naturephysics2015,Qian_ScienceAdvances2018,Halder_PRB2018,Chacon_naturephysics2018,Aqeel_PRL2021,Lee_JPhysCondMat2021}. More quantitative and detailed discussions are available in the next section and SM Section 4 respectively.}

\section*{Reservoir Metrics}

\rev{Unlike software-based reservoirs where their neural numbers/sizes/connections are well-defined by hyperparameters, properties of physical reservoirs cannot be easily mapped onto the corresponding hyperparameters. Here we use task-agnostic reservoir metrics, i.e. nonlinearity (NL), memory-capacity (MC) and complexity (CP)~\cite{dambre_scientificReports2012,Love_arXiv2021} to characterise the reservoir properties (see SM Section 6 for details), and quantitatively discuss the correlation between reservoir performance by normalised MSE (MSE$'$) for different tasks and the metrics.} We performed both forecasting and transformation tasks across a wide range of $H_{\rm c}$ values at 4~K as shown in Fig.~\ref{Fig:4}\revv{h}. In parallel, metric scores are evaluated for each $H_{\rm c}$ as plotted in Fig.~\ref{Fig:4}\revv{i}. 

MSE$'$ for the forecasting tasks is at best in the skyrmion phase and increasingly worse as it enters the conical phase. For transformations, on the other hand, the skyrmion phase exhibits the worst performance compared to the \revv{conical phase}, demonstrating that these trends are clearly correlated with the metrics. In particular, MC shows essentially the same behaviour as MSE$'_{\rm FC}$ with $H_{\rm c}$, suggesting that MC is a key property for better performance in forecasting tasks. \revv{As discussed earlier,} MC in the skyrmion phase stems from the history-dependent fading memory property generated by its gradual skyrmion nucleation with repeated field-cycles~\cite{Aqeel_PRL2021,Lee_JPhysCondMat2021}. As the other phases do not have this property, MC is smaller as it leaves the skyrmion phase. In contrast, rich and complex spin-wave mode dispersion in the conical/ferrimagnetic phases provides the physical basis for high NL and CP scores, offering strong transformation task performance \rev{(see more detailed discussions in Section 4 of SM)}. This highlights the task-adaptive approach and provides examples of how distinct physical phases may be harnessed across a broad range of systems for flexible neuromorphic computing.

The correlation between different parameters can be more visibly identified by the standard Spearman's rank correlation coefficient~\cite{spearmanBook} as shown in Fig.~\ref{Fig:4}\revv{j} \rev{(See SM Section 7 for details)}. Here, the algorithm outputs [-1,~1] where 1 (-1) corresponds to a perfect proportionality (inverse proportionality) \revv{with 0 for no correlation}. Note that since the better performance in each task is represented by lower MSE$'$, the correlation with a negative value to each metric indicates a positive correlation in our analysis. The performance of time series forecasting strongly correlates with MC (-0.89) and CP (0.57), revealing that MC (CP) is favoured (disfavoured) for this particular type of task, while the opposite is true for transformation tasks. It is also important to highlight that MC and CP have a clear negative correlation (-0.68), indicating a trade-off nature between these two reservoir properties. Subsequently, a high correlation between NL and CP (0.67) suggests that a more nonlinear system enhances the amount of meaningful input data encoded in the reservoir, with this benefit offset by a somewhat lower MC as shown by a weak negative correlation between NL and MC (-0.27).

\rev{We} show the specific relationship between reservoir performance evaluated by MSE$'$ and MC (CP) as plotted in Fig.~\ref{Fig:4}\revv{k}~(\ref{Fig:4}\revv{l}), where the colour of the dots encodes which magnetic phase the metrics were evaluated against. See SM \rev{Section 7} for the \rev{plots of} other correlations. Following the Spearman's rank correlation values for each pair, both plots have a negative trend for each reservoir characteristic. Unlike the conical phase, the metrics of the skyrmion phase appear to be clustered in high values of MC between 4 and 7, further confirming that such skyrmion textures are responsible for adding the overall memory to the system for excellent forecasting performance. On the other hand, the system's ability to perform transformation tasks can reach its full potential by maximising the complexity, which occurs when the conical phase dominates the magnet. This sheds light on the importance of the task-adaptive capability of reservoirs when we design and perform multiple tasks by a single physical reservoir device. \rev{Further discussion of the reservoir metrics including their mathematical form and relation to reservoir hyperparameters often evaluated on software-based reservoirs (such as the spectral radius) is provided in SM Section 6.}

\begin{figure}[!h]
\centering
\includegraphics[width=0.9\linewidth]{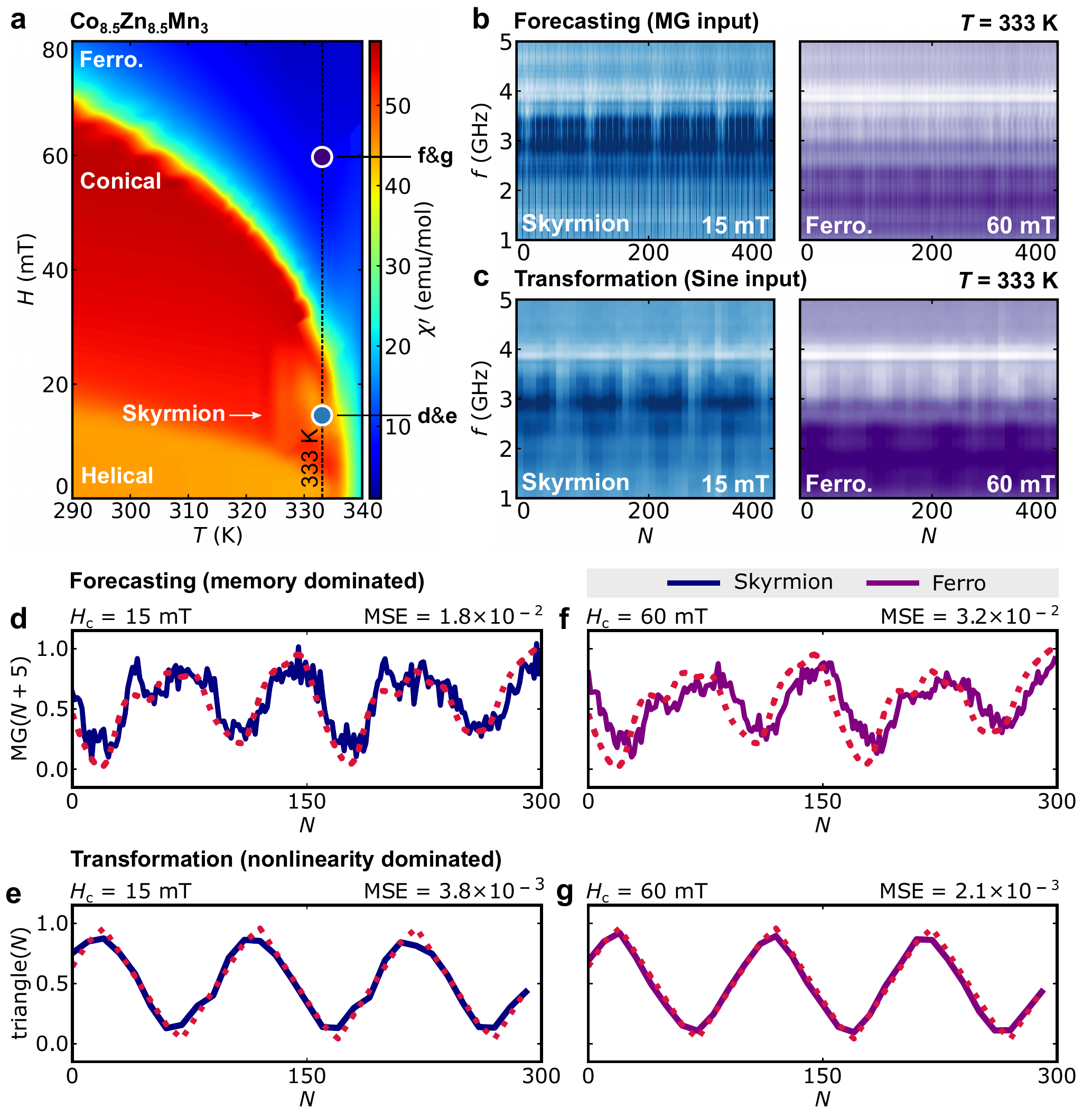} 
\caption{\rev{\textbf{a}, A 2D plot of the real-part of ac susceptibility ($\chi'$) to identify magnetic phases in a Co$_{8.5}$Zn$_{8.5}$Mn$_{3}$ crystal with helical, skyrmion, conical and ferromagnetic phases. The vertical dotted line represents the temperature at which we performed reservoir computing experiments. \textbf{b}, Spin dynamics spectra measured during field cycling $N$ for reservoir computing tasks at different center fields of 15 (left) and 60~mT (right). The top two panels are for the forecasting tasks whereas the bottom two are for transformation, both performed at 333~K. \textbf{c}, Reservoir computing performance of predicting the nonlinear \revv{Mackey-Glass} function for five future steps (top) and transformation from a sine input signal to triangle output function (bottom). The dotted curves/lines are the target function and solid curves/lines are ones generated via our reservoir computing.}}
\label{Fig:5}
\end{figure}

\section*{\rev{Above-room-temperature demonstration}}

\rev{Finally, we present that the task-adaptive reservoir concept can be transferable to different material systems, here using other chiral magnets Co$_{8.5}$Zn$_{8.5}$Mn$_{3}$ (Fig.~\ref{Fig:5}) and FeGe (see SM section 9). Consistent with earlier work of the same class of materials Co-Zn-Mn (e.g. Refs.~\cite{Karube_PRMater2017,Takagi_PRB_2017}), multiple magnetic phases in Co$_{8.5}$Zn$_{8.5}$Mn$_{3}$ can be clearly recognised in a plot of ac susceptibility measurements shown in Fig.~\ref{Fig:5}a. In particular, in the vicinity of its Curie temperature, we can recognise the signature of a thermodynamically stable skyrmion phase - see also Fig.~S8 in SM that shows the imaginary part of the ac susceptibility to highlight this phase. We therefore constructed physical reservoirs by applying our field cycling scheme at 333~K with different magnetic centre fields $H_\text{c}$~=~15 and 60~mT with 10~mT cycling width. In Figs.~\ref{Fig:5}b\&c, we show the spectra of magnetic resonance during field cycling of both nonlinear Mackey-Glass and sine input functions to carry out the future prediction and transformation tasks, respectively. For both tasks, we observe that the spectra strongly depend on the centre field, demonstrating the phase-tunability of physical reservoirs in this material. Using these physical reservoirs with different magnetic phases, we performed both tasks, the results of which are displayed in Figs.~\ref{Fig:5}d-g. For the forecasting task (Figs.~\ref{Fig:5}d\&f), the skyrmion-dominated reservoir ($H_\text{c}$~=~15~mT) outperforms the ferromagnetic reservoir ($H_\text{c}$~=~60~mT), in terms of MSE. In contrast, the ferromagnetic reservoir can yield a better MSE than the skyrmion-dominated one for the transformation task (Figs.~\ref{Fig:5}e\&g). See SM Section 10 for the full phase-tunability of \coznmn{} and FeGe. While there is clear space to improve MSE as well as to make full use of the task-adaptive nature of this material system, this above-room-temperature demonstration can show no fundamental limit of using the task-adaptive concept in a wide variety of materials.}

\section*{Conclusion}
We have demonstrated the substantial benefits of introducing a phase-tunable approach and hence task-adaptability to physical reservoir systems. A single physical reservoir may now be actively reconfigured on-demand for strong performance across a broad range of tasks without the requirement for fabricating additional samples or using entirely different physical systems. This approach invites further development, such as online training and dynamic on-the-fly reservoir reconfiguration for incoming real-time data sets. Moreover, \rev{the phase-tunable approach demonstrated in our study can be transferable to a broad range of physical reservoirs, not only to magnetic materials that host chiral spin textures~\cite{Back_IOP2020,yu2021magnetic,Nayak_Nature2017,Karube_PRMater2017}, but also potentially to non-magnetic systems having rich thermodynamical phase diagrams. It might also offer additional functionality for wave-based physical recurrent neural networks using acoustics~\cite{Hughes_SciAdv2019} and spin-waves~\cite{papp2021nanoscale}.} Experimental demonstration of on-demand reservoir reconfigurability brings physical reservoir computing closer to fully realising its promise and helps develop an alternative to CMOS-powered software neural-network approaches.

\section*{Acknowledgments}
O.L. and H.K. thank the Leverhulme Trust for financial support via RPG-2016-391. W.R.B. and J.C.G. were supported by the Leverhulme Trust (RPG-2017-257) and the Imperial College London President's Excellence Fund for Frontier Research.
J.C.G. was supported by the Royal Academy of Engineering under the Research Fellowship programme. This work was partly supported by Grants-In-Aid for Scientific Research (18H03685, 20H00349, 21K18595, 21H04990, 21H04440, 22H04965) from JSPS, PRESTO (JPMJPR18L5) and CREST (JPMJCR1874) from JST, Katsu Research Encouragement Award and UTEC-UTokyo FSI Research Grant Program of the University of Tokyo, Asahi Glass Foundation.
 This work has also been funded by the Deutsche Forschungsgemeinschaft (DFG, German Research Foundation) under SPP2137 Skyrmionics, TRR80 (From Electronic Correlations to Functionality, Project No. 107745057, Project G9), and the excellence cluster MCQST under Germany's Excellence Strategy EXC-2111 (Project No. 390814868). \revv{We thank Adnan Mehonic, Pavlo Zubko and Antonio Lombardo for reading an earlier version of manuscript and provided comments.}

\section*{Author Contributions}
O.L., K.D.S., J.C.G. and H.K. designed the experiments. 
O.L., T.W. \rev{and D.P.} performed measurements.
O.L., T.W., K.D.S., J.C.G., W.B. and H.K. analysed the results with help from the rest of the co-authors. 
S.S., A.A., \rev{K.K., N.K., Y.Ta.,} C.B. and \rev{Y.To.} \rev{grew the chiral magnetic crystals and characterised magnetic properties of them.} H.K. proposed and supervised the studies. 
O.L., J.C.G. and H.K. wrote the manuscript with inputs from the rest of the co-authors.

\section*{Data availability}

The data presented in the main text and the Supplementary Information are available from the corresponding authors upon reasonable request.

\section*{Code availability}
The code used in this study is available from the corresponding author upon reasonable request.

\bibliography{sn-bibliography}

\end{document}


\flushbottom
\maketitle

\section{Details of experimental setup}
\label{SEC:ExperimentSetup}
\subsection{Ferromagnetic Resonance}

We employ a vector network analyser (VNA; Rohde \& Schwarz ZNB40) to measure the spectral response of \rev{chiral magnetic} crystals via ferromagnetic resonance (FMR) techniques. In our experiment, the sample crystal is placed on a coplanar waveguide (CPW) board which sits on a copper cold finger of a closed-cycle helium-cryostat. \rev{For Cu$_2$OSeO$_3$, we} apply an external magnetic field $\textit{H}$ along the $\langle 100 \rangle$ crystallographic
direction for efficient generation of the low-temperature skyrmions~\cite{Aqeel_PRL2021,Lee_JPhysCondMat2021}. The microwave reflection coefficient $S_{11}$ is recorded by the VNA as a function of microwave frequency to characterise the spectral response for given magnetic fields and temperatures. For our measurements, we sweep the frequency, comprising 1601 frequency points ($M$) at 0~dBm applied microwave power. Thus, a single raw spectral recording of $S_{11}$ consists of 1601 frequency points, which is associated with the frequency dependence of dynamic magnetic susceptibility $\chi_{\rm m}$.

\subsection{Field-cycling scheme}

\begin{figure}[tp]
\centering
\includegraphics[width=1\linewidth]{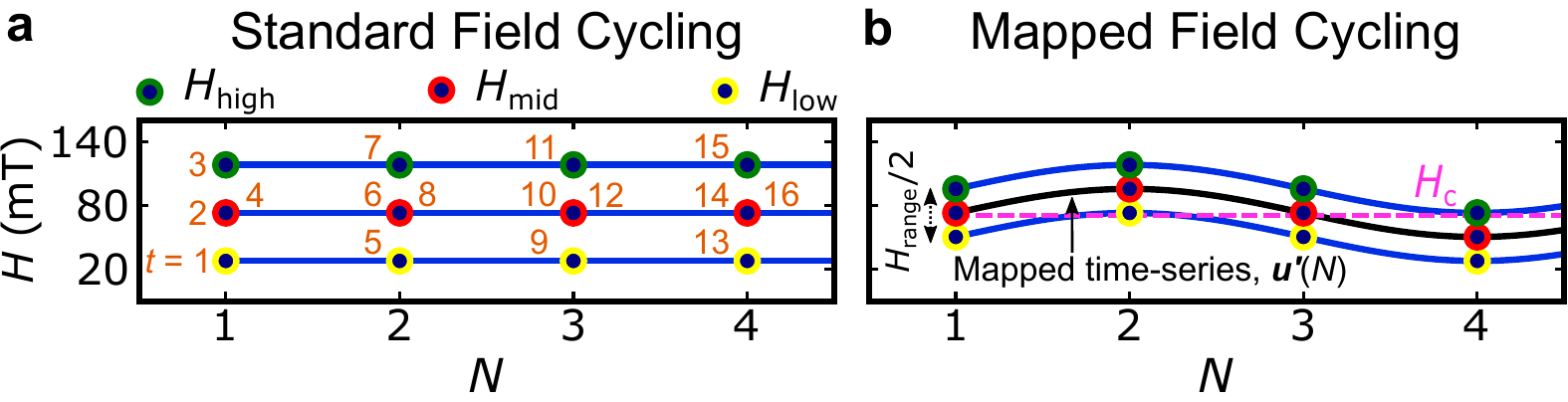}
\caption{\textbf{a\&b}, Comparison of standard field-cycling (\textbf{a}) and mapped field-cycling schemes (\textbf{b}). Typical input scheme as a function of $N$ for $H_{\rm c}$~=~73~mT and $H_{\rm range}$~=~90~mT. Here, $t$ denotes the order of applied field values and measurement points. In standard cycling schemes, the lower and the upper field values are fixed, whereas, in mapped field-cycling, such field values are modulated by the input function at a central field value $H_{\rm c}$ with a fixed separation between the upper ($H_{\rm high})$ and the lower ($H_{\rm low}$) cycling fields, described by $H_{\rm range}\rev{/2}$.}
\label{Fig:MFC}
\end{figure}

In standard field-cycling schemes without envelope modulation, a single field-loop $N$ is completed when $\textit{H}$ is increased and decreased between fixed field points, e.g. defined by $H_{\rm low}$ (yellow), $H_{\rm mid}$ (red) and $H_{\rm high}$ (green) in Fig.~\ref{Fig:MFC}a, having different time steps of $t$ as labelled. During the cycling process, the VNA records the corresponding spectra for each magnetic field value to study the nucleation of metastable lattices such as low-temperature skyrmions~\cite{Aqeel_PRL2021,Lee_JPhysCondMat2021}. This cycling scheme, however, lacks the ability to construct a time series input function for reservoir computing.

We have therefore established the mapped field-cycling (MFC) scheme to apply the field-cycling data input protocol for physical reservoir computing. This technique, as shown in Fig.~\ref{Fig:MFC}b, modulates each of $H_{\rm low}$, $H_{\rm mid}$ and $H_{\rm high}$ for different $t$ to generate a field-cycling-dependent input function $\textbf{\textit{u'}}(N)$. This makes it possible to incorporate arbitrary time series signals $\textbf{\textit{u}}(t)$ in our scheme. For the mapping procedure, $\textbf{\textit{u}}(t)$ is normalised between [-1, 1] and offset by a central cycling field value $H_{\rm c}$, where two additional copies ($H_{\rm high}$ and $H_{\rm low}$) are generated above and below $H_{\rm mid}$ using the cycling width $H_{\rm range}$. In this work, we accommodated two specific input sequences to suit different target applications: a chaotic oscillatory Mackey-Glass time series signal~\cite{Mackey_1977} for forecasting and a sinewave for transformations. We construct the reservoir outputs using the FMR spectra measured at the lowest field point (yellow dots) within the cycles.

\rev{The MFC scheme therefore allows FMR frequency-multiplexing. Frequency-multiplexing is a technique commonly used to broadcast a single-dimensional input signal to multiple outputs. In our experiments, each field point is encoded as a series of frequencies applied to the magnetic system. By measuring the FMR response, multiple output signals at different frequencies can be separated and analysed in the spectral space to be used for computation.}

\rev{A typical time to solve tasks is in total around two hours. The breakdown of this entire process is: 1. input field mapping as pre-processing (less than one minute), 2. inputting data as magnetic field and recording physical reservoir output via VNA (2 hours) and 3. training/testing reservoirs (less than 1 minute). For the reservoir construction process, we use a VNA to acquire frequency spectra, which take approximately one second per single spectrum. Changing the magnetic field dominates the measurement time, and the timescale is limited by this speed.}

\begin{figure}[t]
\centering
\includegraphics[width=1\linewidth]{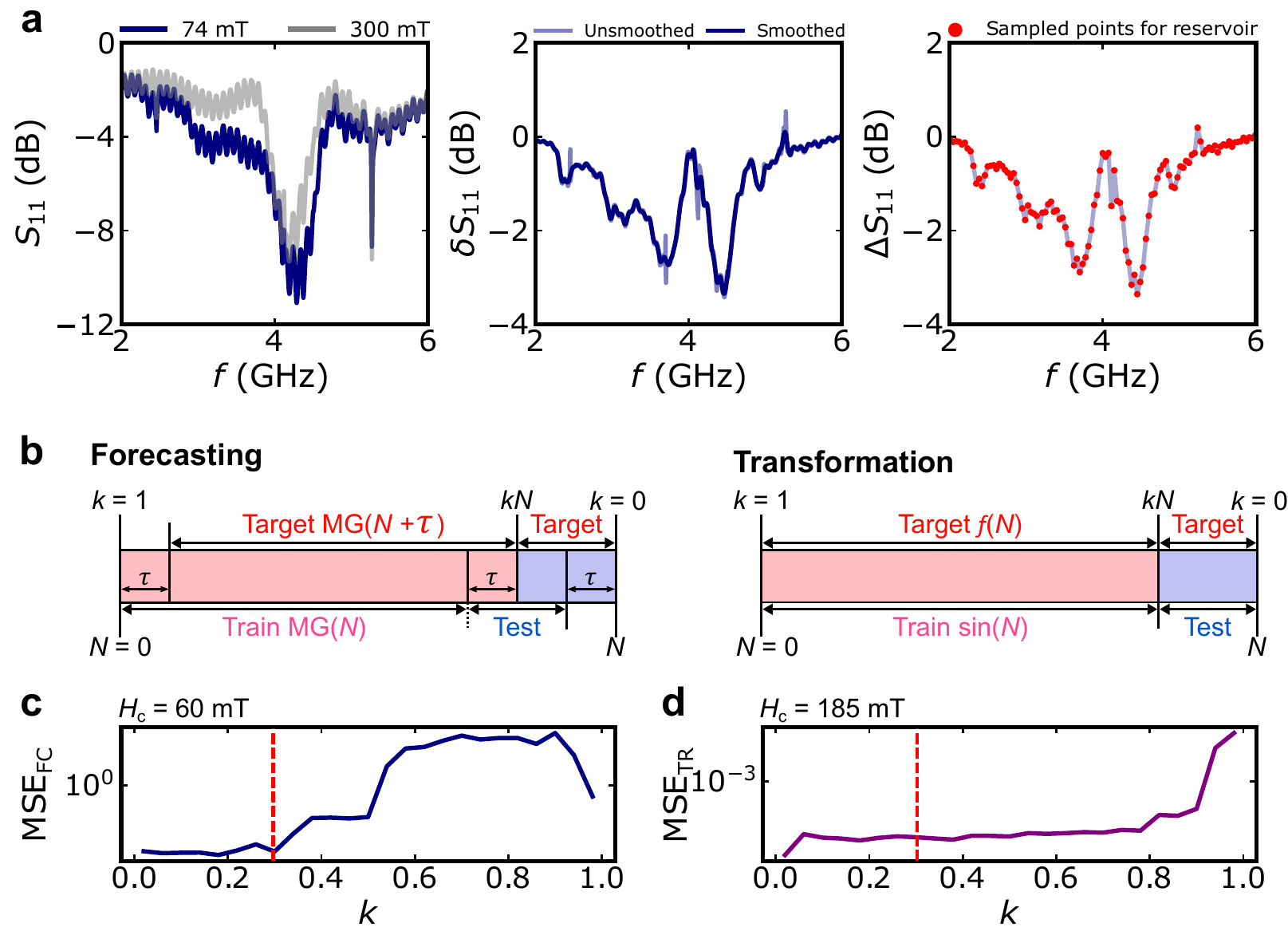}
\caption{\textbf{a}, Definition of $\Delta S_{11}$; a smoothed spectra ($\delta S_{11}$) is sampled at fixed intervals after subtracting high-field spectra from the raw signal ($S_{11}$). \textbf{b}, Illustration of training and testing dataset. For forecasting, the weights are calculated on the training readouts of the reservoir constructed using a Mackey-Glass input MG(\textit{N}) to predict the target MG($N+\tau$), defined at $\tau$ future steps from the original. For transformations, the reservoir is created using a sinewave input sin(\textit{N}), where the weights are calculated on the training data to best reproduce a target function $f(N)$ that is different to a sinewave. Here, $k$ is a constant multiplier between [0, 1] that determines the length of the test set data. \textbf{c}\&\textbf{d}, MSEs of forecasting and  transformations as a function of $k$. The red dotted line highlights the chosen value of $k$ used in this study (0.3).}
\label{Fig:DataProcessing}
\end{figure}

\section{Details of reservoir computing protocols}
\label{SEC:RC_protocol}

\revv{\subsection*{Data processing}}

After completing a set of MFC measurements, the spectra data are pre-processed before being added to the reservoir matrix $\textbf{\textit{R}}$ as shown by an example in Fig.~\ref{Fig:DataProcessing}a. Each spectrum undergoes the same processing method of a high-field (300~mT) background subtraction, a numerical lossless smoothing accommodated by the Savitzky-Golay filter~\cite{savitzky_golay_1964} and a spectrum sampling at fixed intervals. Data sampling is necessary to avoid an over-fitting problem caused by too many data points during training \rev{(see Section~\ref{SEC:SpectraFeatures} and Fig.~\ref{Fig:SpectralSize} for more details)}. The sampling interval is determined by an automated search process that best produces the mean squared error (MSE) of the test data.

\revv{\subsection*{Target generation}}

\revv{The transformation targets shown in the main text have been generated with \texttt{scipy.signal} package~\cite{2020SciPy}, where the input array is defined by $0.2\pi \left \{1..N\right \}$ for the `square' waveform with a duty cycle of 0.5 and a `saw' signal with a width of the rising ramp as 1. Note that the square target waveform has a very slight slope between high and low values due to the finite sample rate.}

\revv{For forecasting tasks, Mackey-Glass, a chaotic time series derived from a nonlinear time-delayed differential equation, was employed. Its complex behaviour is commonly used as a benchmark for testing the performance of prediction algorithms. The signal is defined by: $\frac{\mathrm{d}x }{\mathrm{d}t} = \beta \frac{x_{d}}{1+x_{n}^{d}}-dx$. We have numerically generated the signal with the following parameters: $\beta=0.2$, $n=10$, and $d=17$.}

\revv{\subsection*{Training and testing}}

For training and testing, $\textbf{\textit{R}}$ is subsequently separated into training and test datasets determined by a test-length factor $k$, which ranges between 0 and 1, as illustrated in Fig.~\ref{Fig:DataProcessing}b. The training dataset is passed on to a variant of the linear regression algorithm\rev{,} ridge regression~\cite{hilt_seegrist_1977} to calculate the \rev{optimal} weights to reproduce the target dataset. \rev{Ridge regression is a common regression technique with a regularisation term $\alpha$ for analysing multicollinear data. The weights are determined by: $\min(w))\left | \right | \chi w-y \left | \right |_{2}^{2}+\alpha \left | \right |w\left | \right |_{2}^{2}$, where $w$, $\chi$, and $y$ are the ridge coefficients (weights), reservoir elements, and the target value terms, respectively. Here, $\alpha$ helps penalise large $w$ obtained during the fitting process to stabilise the model and prevent overfitting. We have used the \texttt{scikit-learn} package for this calculation~\cite{scikit_learn}.} The obtained weights are then applied to the unseen test dataset to evaluate the training performance and compared to the test target data. 

In this study, $k$ was fixed at 0.3, thus using 70~\% of data for training and 30~\% for testing with $N$~=~1000 cycles. The dependence of $k$ on the MSE values for two different tasks is shown in Fig.~\ref{Fig:DataProcessing}c. As can be observed in these plots, sufficient training data are necessary to improve MSE for each case; in other words, $k$ should be reasonably smaller than the unity. We confirm that the choice of $k$ does not significantly alter our analysis and conclusions drawn in this study.

\revv{\subsection*{Performance evaluation of RC}}

\rev{MSE is a statistical measure that quantifies the difference between predicted and true values by averaging the squared differences across data points. A lower MSE value indicates a better predictive performance for a given task. We calculate our MSE values using the \texttt{mean\_squared\_error} function from the \texttt{sklearn.metrics} package~\cite{scikit_learn}, which evaluates: ${\rm MSE}(y,\hat{y})~=~\frac{1}{n_{\rm samples}}\sum_{i=0}^{n_{\rm samples}-1}(y_i-\hat{y_i})^2$, where $y_i$ and $\hat{y_i}$ corresponds to the target (true signal) and transformed/predicted values, respectively, and each consists $n_{\rm samples}$ number of data points.}

\section{\rev{Magnetic phases and tunability in Cu$_2$OSeO$_3$}}
\label{SEC:PhaseTuanability}

Cu$_2$OSeO$_3$ \rev{is one of chiral magnets, having} a noncentrosymmetric cubic lattice belonging to the P2$_1$3 space group where the competition of symmetric/anti-symmetric exchange, magnetic-dipole and Zeeman interactions provides different magnetic phases~\cite{seki_science2012, Onose_PRL2012}. \rev{As shown in Fig.~2a in the main manuscript, it possesses four different magnetic phases in the temperature-magnetic-field phase diagram. For small magnetic fields, spins in Cu$^{2+}$ ions point as spiral rotation within a specific plane, hence having the corresponding modulation vector. This is called the helical phase. When the magnetic field is increased, there \revv{is} a finite spin component along the field direction for each Cu$^{2+}$ ion, forming the conical state. Finally, when the magnetic field is further increased, the spiral component is completely lost, having the three-up/one-down spin configuration. This is called the ferrimagnetic state~\cite{seki_science2012}. Furthermore}, skyrmion phases tend to form between the helical and conical phases at high-temperature pockets closely below the Curie temperature $T_{\rm c}$~\cite{seki_science2012,Onose_PRL2012}. However, a distinctive thermodynamically metastable skyrmions can also be realised at lower temperatures~\cite{Chacon_naturephysics2018,Halder_PRB2018}. Their population can be controlled by the number of field-cycling~\cite{Aqeel_PRL2021,Lee_JPhysCondMat2021}, making them an adequate candidate to perform reservoir computing. From a detailed previous study~\cite{Lee_JPhysCondMat2021}, the sample is anticipated to host this phase at temperatures below $\sim$25~K and magnetic fields between 25~$<\textit{H}<$~120~mT.

\begin{figure}[h]
\centering
\includegraphics[width=1\linewidth]{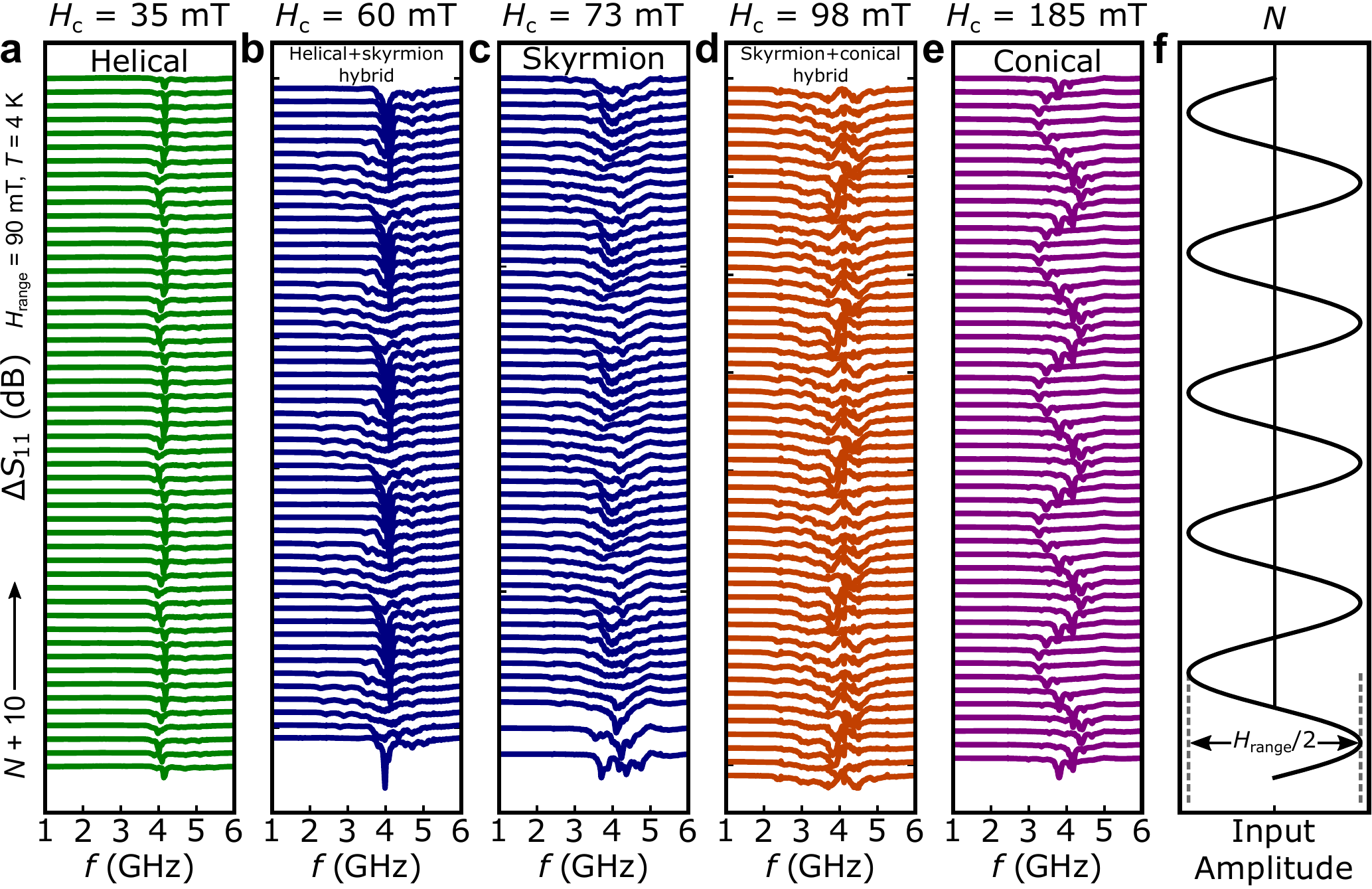}
\caption{\rev{Evolution of spectra in Cu$_2$OSeO$_3$ as a function of microwave frequency $f$ for increasing values of $N$ for a sinewave input signal at different values of $H_{\rm c}$ with $H_{\rm range}$~=~90~mT at a constant temperature $T$~=~4~K. Each spectrum represents the completion of 10 field-cycles from its previous. By tuning $H_{\rm c}$, the reservoir can be constructed with different dominant magnetic phase-spaces: \textbf{a}, helical, \textbf{b}\revv{, helical~+~skyrmion hybrid,} \textbf{c}, skyrmion, \textbf{d}, hybrid (skyrmion~+~conical), and \textbf{e}, conical. \textbf{f}, Applied input amplitude as a function of $N$ used to construct reservoirs in \textbf{a}-\textbf{e}.}}
\label{Fig:SpectraWaterfall}
\end{figure}

Figure~\ref{Fig:SpectraWaterfall} summarises the spectral evolution by a sinewave input signal for different $H_{\rm c}$. These plots use the same dataset that generate colour plots of Fig.~2e in the main manuscript. For $H_{\rm c}$~=~35~mT, the magnet is predominated by the helical phase (Fig.~\ref{Fig:SpectraWaterfall}a) \revv{and its resonant modes around 4 GHz are assigned as $\pm Q$ modes in the magnetic phase~\cite{Garst_JPhysD2017,Lee_JPhysCondMat2021,Aqeel_PRL2021}}. However, increasing $H_{\rm c}$ to 60~mT and 70~mT (Figs.~\ref{Fig:SpectraWaterfall}b\&c), clear signatures of skyrmions are seen (between 2~-~3~GHz), where their FMR positions are subtly modulated by the input signal. At $H_{\rm c}$~=~98~mT (Fig.~\ref{Fig:SpectraWaterfall}d), approximately half of the input signal is cycled within the skyrmion phase and the other half in the conical phase. This results in a hybridised state where both skyrmion and conical modes share the lattice. Furthermore, when field-cycling predominantly occurs outside of the skyrmion phase, i.e., $H_{\rm c}$~=~185~mT (Fig.~\ref{Fig:SpectraWaterfall}e), the conical reservoir excellently encodes the input signal as shown by their FMR positions\rev{, comparable to the amplitude of the input fields as depicted in Fig.~\ref{Fig:SpectraWaterfall}. This mode yields} a high \rev{nonlinearity (NL) and complexity (CP)}, leading to an outstanding performance of transformation tasks.

\section{Computational properties associated with physical characteristics}
\label{SEC:SpectraFeatures}

\begin{figure*}[b!]
\centering
\includegraphics[width=1.0\linewidth]{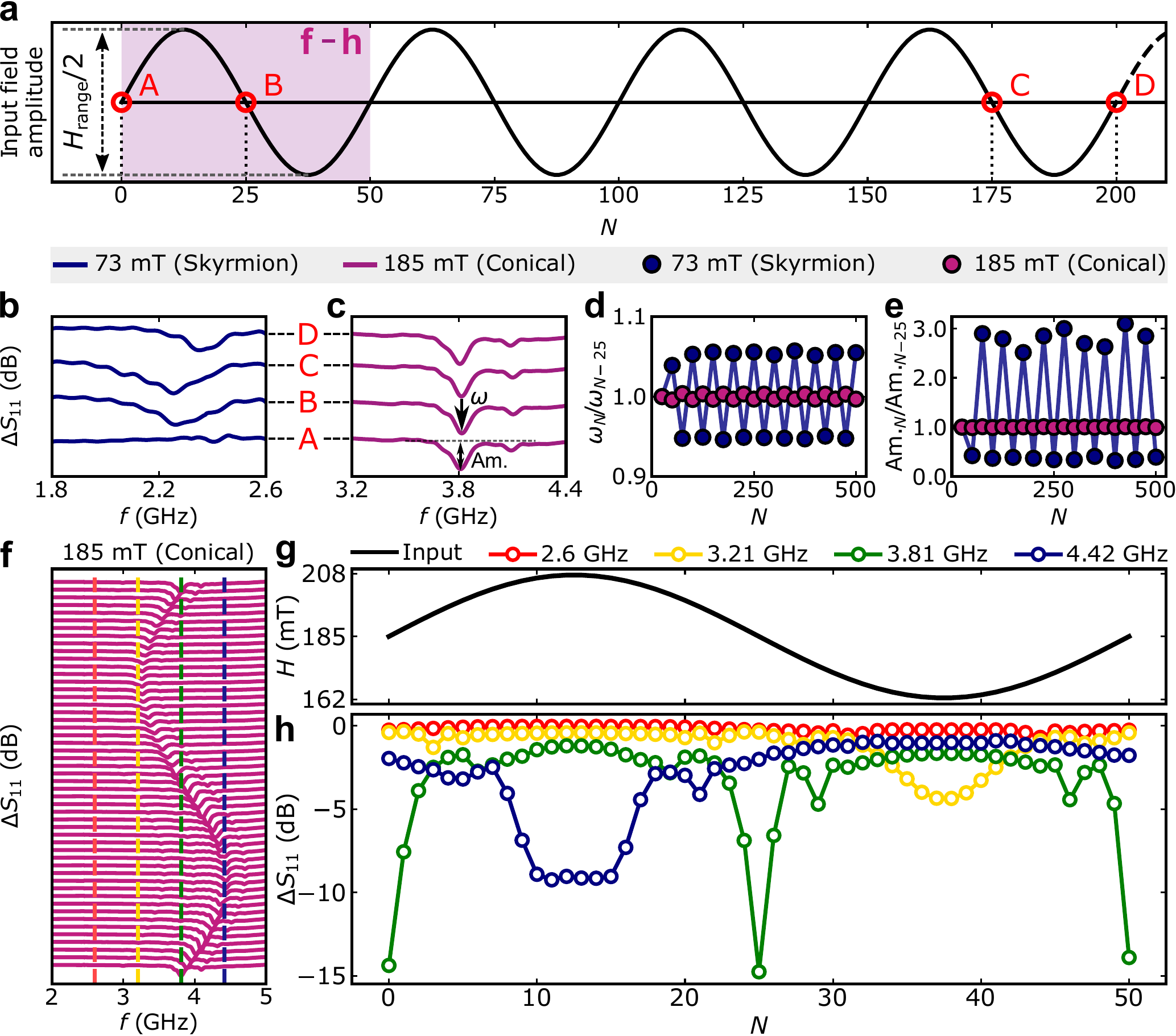} 
\caption{\rev{Spectral evolution and feature analysis of physical reservoirs using Cu$_2$OSeO$_3$. \textbf{a}, Illustration of the MFC scheme for defining input amplitudes, with points marked ``A" to ``D" representing input fields analyzed in \textbf{b}-\textbf{e}. The purple shaded region highlights a single period of sinewave used for analysis in \textbf{f}-\textbf{h}. \textbf{b}\&\textbf{c}, Individual spectra highlighting counterclockwise skyrmion and conical modes, respectively. \textbf{d}\&\textbf{e}, Relative changes in peak frequency ($\omega_{N}/\omega_{N-25}$) and peak amplitude ($\rm{Am}._{\mathit{N}}/\rm{Am}._{\mathit{N}-25}$), measured at each node on the sinewave input, as an evolution of $N$. Dots and lines in blue (purple) represent the skyrmion (conical) mode. \textbf{f}, Cycling-evolution of spectra at conical phase ($H_c=185$~mT) as a function of frequency. The dashed vertical lines depict the frequencies evaluated for \textbf{h}. \textbf{g}, First period of sinewave input fields as a function of $N$. \textbf{h}, Spectral amplitudes as a function of cycling number $N$ at various fixed frequency channels.}}
\label{Fig:SpectraFeatures}
\end{figure*}

\rev{Here we further analyse results in Fig.~\ref{Fig:SpectraWaterfall} to describe our interpretation of why individual magnetic phases perform differently and are suited for the specific tasks we present. We show in Fig.~\ref{Fig:SpectraFeatures}, our further analysis of each spectrum recorded at the exact field strengths (73 and 185~mT) but at different points. Figure~\ref{Fig:SpectraFeatures}a summarises our sine input function field series with cycle number $N$ and specifies Points A~-~D all having the same field value but different $N$, i.e. A ($N=0$), B ($N=25$), C ($N=175$) and D ($N=200$), respectively. For the spectral evolution at the centre field of 73~mT targeting at the skyrmion modes in Cu$_2$OSeO$_3$ (Fig.~\ref{Fig:SpectraFeatures}b), all of the frequency spectra (A~-~D) are dissimilar to each other, although they are measured at the exact same magnetic field value. For example, the spectra at A (C) and B (D) are separated by the half period of the sine input function and this field history is imprinted as their magnetic properties, i.e., its absorption properties $\Delta S_\text{11}$. This clearly shows short-term memory capacity (MC) in the skyrmion phase: $i.e.$ $x(N) = f(..., u'(N-1), u'(N), x(0))$ where $x(N)$ is the state vector of a reservoir at the field cycling point $N$ and $u'(N)$ is the input function at $N$.} 

\rev{Moreover, Points B and C are connected by periodic translation, and their spectra have the same peak position but different heights. This is due to the cycling-number-dependent meta-stable skyrmion population – the more we cycle, the more we nucleate the skyrmions~\cite{Aqeel_PRL2021,Lee_JPhysCondMat2021}. This intrinsic material property generates additional (long-term) memory in the reservoir, being able to perform superbly for tasks requiring strong MC, such as future prediction. High performance (MSE) of the skyrmion phase on our benchmarking of predicting the nonlinear Mackey-Glass time series function, correlated with high values of MC, is robust evidence of this claim. In this task, the relationship between input data excitation and target output response (i.e. prediction value) is constantly evolving throughout the Mackey-Glass time series due to its chaotic nature. In order to reliably predict these ever-changing targets, the state of the reservoir must hold enough information about past states (the short-term memory) to accurately discriminate the precise nature of the chaotic behaviour at the current position in the time series.}

\rev{This unique memory property is absent for reservoirs dominated by the other two magnetic phases as observed in Fig.~\ref{Fig:SpectraFeatures}c for the conical reservoir where spectra for Points A~-~D are all identical, therefore history-independent. To quantitatively discuss the difference in the reservoir memory property, we performed a single Lorentzian fit~\footnote{\rev{A curve-fitting technique commonly used to extract physical quantities from the FMR response. Defined by: $\Delta S_{11} \propto (\rm{Am.} \Delta \omega)/((\mathit{f} - \omega)^2+\Delta \omega^2 + \rm{C})$, where Am., $\Delta \omega$, $f$, $\omega$, and C, represents the peak amplitude, linewidth, frequency, peak frequency position, and offset constant, respectively.}} to each spectrum shown in Figs.~\ref{Fig:SpectraFeatures}b\&c for extracting their peak position ($\omega_N$) and amplitude (Am.$_N$). We plot the ratio of $\omega_N$/$\omega_{N-25}$ and Am.$_N$/Am.$_{N-25}$ for both skyrmion and conical reservoirs in Figs.~\ref{Fig:SpectraFeatures}d\&e. Both plots clearly support the strong memory property in the skyrmion reservoir.}

\rev{Instead, the conical reservoir is equipped with high NL and CP, yielding strong performance in the transformation tasks that require these properties. To feature the strong NL/CP property of the conical mode, we plot individual spectra for one sine input function period (Fig.~\ref{Fig:SpectraFeatures}f) together with its spectral value evolution with $N$ in Fig.~\ref{Fig:SpectraFeatures}h. The main peak of the conical phase moves in a similar manner as the input magnetic field, as demonstrated in Fig.~\ref{Fig:SpectraFeatures}f. Individual frequency values plotted in Fig.~\ref{Fig:SpectraFeatures}h have drastic changes in amplitude, distinctly different from the sine input function plotted above (Fig.~\ref{Fig:SpectraFeatures}g). Each frequency point has unique evolution offering rich nonlinear responses as a whole. This large set of diverse responses to the input function empowers the reservoirs in performing signal transformation tasks, as numerically quantified in our metrics, i.e. NL and CP.}

\begin{figure*}
\centering
\includegraphics[width=1\linewidth]{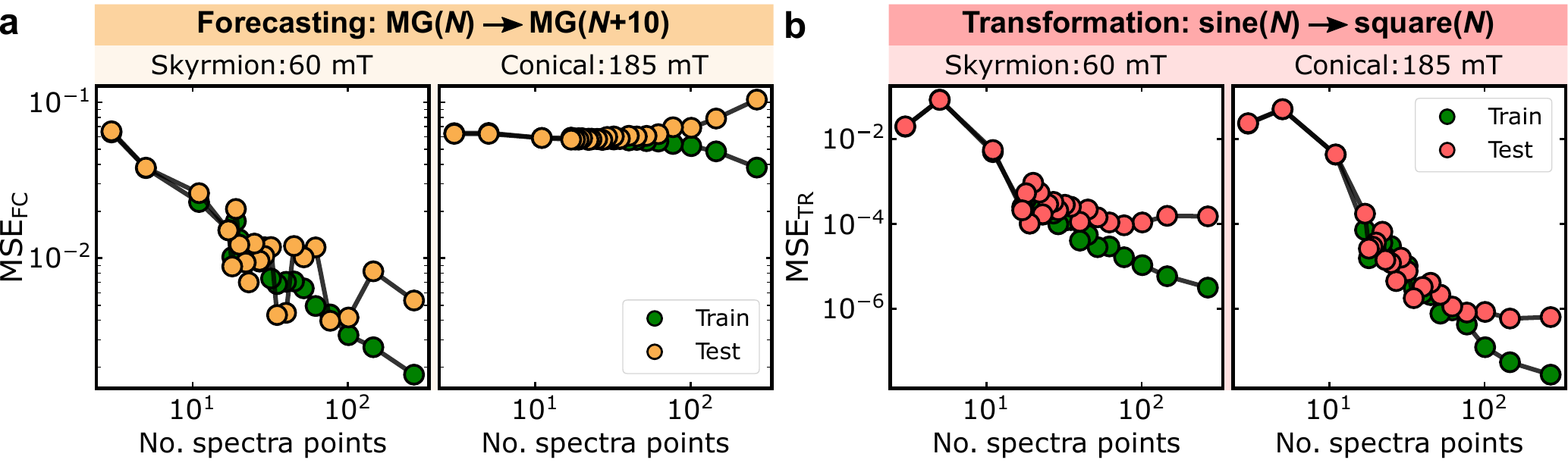} 
\caption{\rev{\textbf{a\&b}, Spectral size-dependence of MSE values for forecasting (MG($N$~+~10); \textbf{a}) and nonlinear transformation (sine to square; \textbf{b}). The left column depicts a skyrmion-dominated region (60~mT), while the right column represents a conical mode-dominated region (185~mT). Green dots denote the training data and yellow (red) dots represent test results for the forecasting (transformation) tasks.}}
\label{Fig:SpectralSize}
\end{figure*}

\rev{Furthermore, the size of reservoirs, i.e. the number of spectral points used for developing our reservoir, is found to be critical in our case. As shown in Fig.~\ref{Fig:SpectralSize}, MSE is greatly improved when the spectral point is increased, except for the future prediction task with the conical reservoir where no memory properties are expected, suggesting that adding more spectral points with no memory does not improve computing performance significantly. Since each spectral point behaves differently by the input function of the magnetic field due to nonlinearity, this high-dimension mapping is very efficient, producing an excellent performance for the benchmarking tasks used in our study.}

\revv{\section{Additional reservoir computing performance}}
\label{SEC:Target_generation}

\begin{figure}
\centering
\includegraphics[width=1\linewidth]{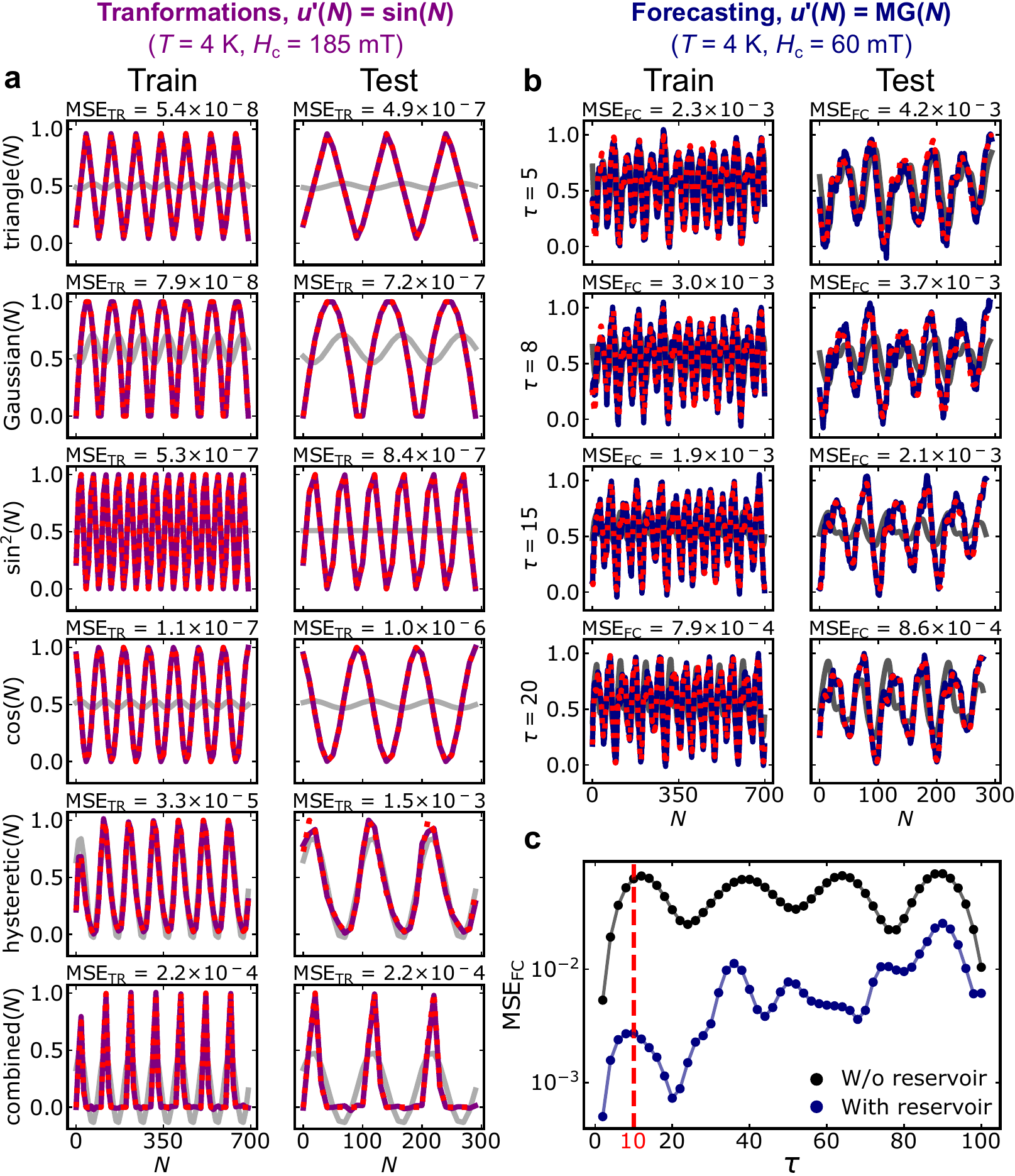}
\caption{\textbf{a}\&\textbf{b}, Training and testing transformations (\textbf{a}) of a sinewave input signal into various target functions with $H_{\rm c}$~=~185~mT (conical), and forecasting (\textbf{b}) of a Mackey-Glass signal for different future steps $\tau$ with $H_{\rm c}$~=~60~mT (skyrmion). Both tasks use $H_{\rm range}$~=~90~mT at $T$~=~4~K. The grey lines are transformation/prediction computation without using the reservoir element. \textbf{c}, Forecasting MSE as a function of $\tau$ using the same conditions as \textbf{b} with and without reservoir. The prediction performance without the reservoir shows an intrinsic periodicity of $\tau\approx$~22, where a red line is drawn at $\tau=10$, representing the presented and the analysed value of $\tau$ in the main text.}
\label{Fig:RCPerformance}
\end{figure}

Next, we present additional reservoir computation results on a broader range of tasks to further support our claims in the main manuscript. Figure~\ref{Fig:RCPerformance}a shows the training and testing performance for nonlinear transformation from a sinewave input signal to a range of different target waveforms. The target data for the triangle and Gaussian signals are generated using the \texttt{scipy.signal} package~\cite{2020SciPy}, similar to the square target waveform. Note that a triangle signal is a form of a saw signal with a symmetric width of 0.5. A periodic Gaussian pulse is generated by concatenating Gaussian signals with a standard deviation of 5. A sine-squared target is constructed by squaring a numerically evaluated sinewave with the same input array used for the square. A cosine waveform has been generated in a similar fashion. A hysteretic signal is a form of a second-order nonlinear equation where the output is dependent on its previous value~\cite{Gartside_naturenano2022}. A combined signal is arbitrarily generated by multiplying a square waveform by all signals shown in the figure ($\rm{square}(\textit{N})\times ... \times \rm{hysteretic}(\textit{N})$). In all cases, the transformation MSE (MSE$_{\rm TR}$) values are within the magnitudes of $10^{-3}$ or below, showing excellent transformation performance across a diverse range of target signals.

Figure~\ref{Fig:RCPerformance}b shows the training and testing data for a forecasting task of the chaotic Mackey-Glass input signal MG(\textit{N}) for a range of different future steps $\tau$. As shown in Fig.~\ref{Fig:RCPerformance}c, although the Mackey-Glass signal is chaotic, it is also quasi-periodic with an approximate period of $\tau\approx22$ (see the black dots). Thus prediction performance (MSE$_{\rm FC}$) also varies periodically, as predicting a full period ahead requires far less modification of the input signal to accurately recreate the target signal, as evidenced by the observed periodic relation between MSE$_{\rm FC}$ and $\tau$. Throughout this study, we have consistently chosen $\tau=10$ where the computation of forecasting is nontrivial within the first quasi-period in order to best evaluate the computational power of our physical reservoir system.

\section{Details of reservoir metric calculations}
\label{SEC:RC_metrics}

We evaluate our reservoir metrics \rev{(NL, MC, and CP)} based on prior works described in \revv{refs.}~\cite{Jaeger_shortterm2002,Roy_2007,dambre_scientificReports2012,Love_arXiv2021}. The underlying scope of these metrics is to quantify how the reservoir states $\chi$($N$) responds to a given input signal, \textbf{\textit{u’}}(\textit{N}). While a random input dataset is often used for such metric calculations, here, we perform metric calculations on the output data from the chaotic Mackey-Glass dataset as the input, as employed in prior studies~\cite{Gartside_naturenano2022}. We use \textbf{\textit{u’}}(\textit{N})~=~MG(\textit{N}) at $H_{\rm range}$~=~90~mT for different values of $H_{\rm c}$ and temperatures to measure the metrics in this work. Before the evaluation, reservoir outputs which are unresponsive to the input signal (i.e. FMR frequency channels in `dead' frequency ranges where no FMR resonance modes are present) and potentially introduce artefacts in reservoir metric calculations are removed. This is achieved by ranking the FMR frequency output channels by their range (maximum value minus minimum value of that frequency channel over the entire set of $N$) and selecting the top 30\% of FMR frequency channels with the highest range to perform our reservoir metric calculations. Channels with low range are effectively noise dominated `dead' regions of the spectra and do not contribute meaningful reservoir metric information.

NL returns a score between [0, 1], where the value of 0 (1) represents a completely linear (nonlinear) system~\cite{Love_arXiv2021}. This metric evaluates the ability of the reservoir to predict the true readout $\chi_{\rm m}$ (FMR frequency channels) when shown up to 8 previous inputs, as shown in Eq.~S1. 
\begin{equation}
{\hat{\chi_m}(N) = \sum_{i=0}^{8}w_iu'\left(N-i\right)}
\end{equation}
Here, the weights $w$ are calculated using linear regression with 750 cycles for training. Subsequently, $w$ is applied to the unseen 250 cycles to obtain the prediction $\hat{\chi}_{\rm m}$, which is compared with corresponding $\chi_{\rm m}$ through $\rm R^2$ coefficient \rev{of determination (see Eq.~9 in Ref.~\cite{Love_arXiv2021} for details)}. This process is repeated for all values of FMR frequency channels ($m$~=~\{1..480\}), where \rev{NL} is evaluated as:
\begin{equation}
{\rm {NL} = 1 - \rm{AVERAGE}(\rm R^2[\hat{\chi}_{\rm m}, \chi_{\rm m}])}
\end{equation}
MC determines the reservoir's capacity to remember previous information about its inputs, i.e. how much past input information is present in the current readout datapoint~\cite{Jaeger_shortterm2002,Love_arXiv2021}. The metric evaluates the performance of the current state of the reservoir ($\chi_{m}$) in predicting up to its last 8 input states, as described in Eq.~S3. 
\begin{equation}
{\hat{u}(N-i)=\sum_{m=1}^{480}w_{m}\chi_m(N)}
\end{equation}
Here, we obtain $w$ similar to the NL evaluation. However, we use the readout ($\chi_{m}$) to predict the states of the previous input signals. This prediction is subsequently compared with the true state of the input via $\rm R^2$ metric, where MC is evaluated (Eq.~S4).
\begin{equation}
{{\rm MC} = \sum_{i=1}^{8}{\rm R^2}\left[ \hat{u}(N-i), u(N-i) \right ]}
\end{equation}
A high MC score indicates that the reservoir retains a substantial amount of past input data in its current spectral information over a more extended period of past inputs. 

CP determines the effective latent space of the reservoir, i.e., the amount of meaningful information encoded in the spectra~\cite{Love_arXiv2021}. Here, only the readouts of the reservoir are considered for the calculations. We first prepare the readouts into two square matrices, each of dimensions $480\times480$ to calculate the effective rank of the individual matrices~\cite{Roy_2007,Love_arXiv2021}, measuring the exponent of a Shannon entropy of normalised singular vector values evaluated using a singular value decomposition technique~\cite{Shannon_1948}. Subsequently, the average of the two effective ranks gives the CP score. Higher CP values indicate that the system is more perceptive to salient features in the input data.

\rev{In software reservoir computing, additional hyperparameter metrics including the spectral radius may be directly calculated via the matrix of internal reservoir weights (i.e. the fixed, randomised internal structure of the reservoir itself as opposed to the task-specific training weights produced via linear regression). While this is not possible in physical neuromorphic computing systems, both as the internal reservoir structure shifts dynamically in response to input stimuli and as the internal reservoir structure is extremely challenging to fully quantify, the higher-level metrics (MC/NL/CP) which may be more readily assessed from the reservoir response have been shown to be strongly correlated with the internal reservoir hyperparameters. A large spectral radius correlates with strong nonlinearity, and a small spectral radius correlates with strong memory~\cite{Butcher_NeuNet2013,Jaeger_GMDRepoert2002}. Hence, the ability of our phase-tunable approach to reconfigure MC, NL and CP metrics can be seen as evidence that our methodology is capable of dynamically reconfiguring reservoir hyperparameters such as the spectral radius and accordingly the internal reservoir connectivity and structure.}

\begin{figure}
\centering
\includegraphics[width=1\linewidth]{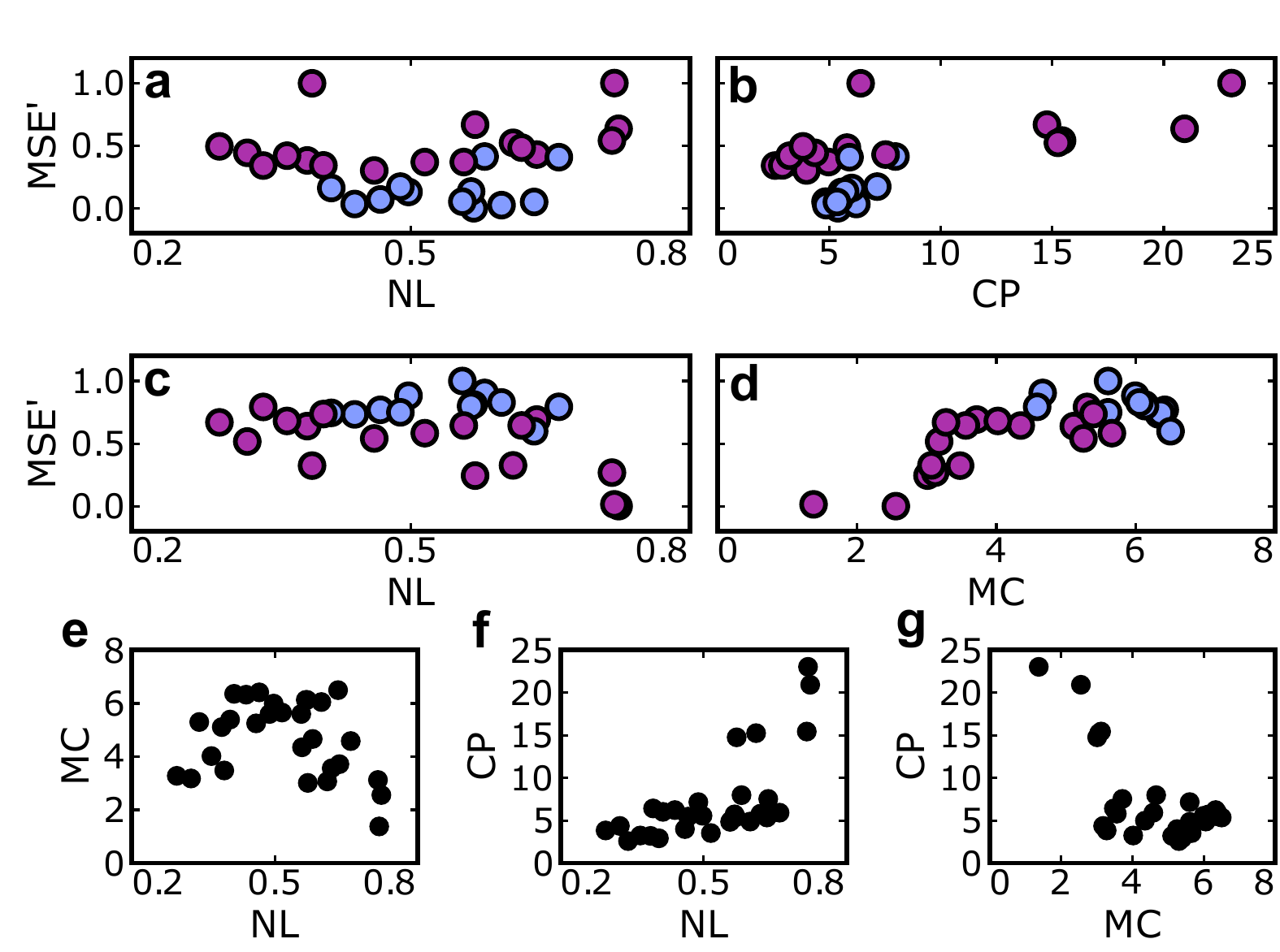}
\caption{\rev{\textbf{a}\&\textbf{b}, Forecasting performance as a function of NL (\textbf{a}) and CP (\textbf{b}). \textbf{c}\&\textbf{d}, Transformation performance as a function of NL (\textbf{c}) and MC (\textbf{d}). \textbf{e}, MC as a function of NL. \textbf{f}\&\textbf{g}, CP as a function of NL (\textbf{f}) and MC (\textbf{g}).}}
\label{Fig:FurtherMetrics}
\end{figure}

\section{\rev{Correlation analysis and additional data}}
\label{SEC:AddtionalCorrelation}

\rev{We determine the correlations of MSE-metric and metric-metrics using the Spearman's rank correlation coefficient~\cite{spearmanBook}, which is a nonparametric (i.e., does not assume the data follows a specific statistical distribution such as the normal distribution) measure of the strength and direction of association between two variables. It reflects the degree to which their rankings correlate, yielding values ranging from -1 to 1. -1 indicates a perfect negative association (where one variable increases, the other decreases). Conversely, 1 implies a perfect positive association (where one variable increases, the other decreases).}

Here we present correlation plots which we do not show in the main text. For this analysis, we normalise the MSE values as follows:
\begin{equation}
{\rm MSE' = \frac{log_{10}(MSE) - min(log_{10}(MSE))}{max(log_{10}(MSE)) - min(log_{10}(MSE))}}
\end{equation}
Note that a log value of MSE was taken to minimise the correlation anomalies arising from a large range of MSE values, resulting in an incorrect representation of the dataset. This is equivalent to plotting the MSE values on a logarithmic scale.

Figures~\ref{Fig:FurtherMetrics}a and b respectively show the MSE$'_{\rm FC}$ as a function of NL and CP. MSE$'_{\rm FC}$ and NL have a weak correlation of 0.14, whereas CP shows a positive correlation of 0.57. MSE$'_{\rm FC}$ is minimised when skyrmions (blue dots) are present. Similarly, Figs.~\ref{Fig:FurtherMetrics}c and d respectively show the transformation performance as a function of NL and MC. MSE$'_{\rm TR}$ negatively correlates with NL (-0.36) and has a strong (positive) correlation with the MC (0.77). In Figs.~\ref{Fig:FurtherMetrics}e-g, we show MC against NL (e), CP against NL (f) and CP against MC (g). In Fig.~\ref{Fig:FurtherMetrics}e, MC is shown as a function of NL, with a relatively low correlation (-0.27) and bell-curve like shape. Fig.~\ref{Fig:FurtherMetrics}f shows the CP as a function of NL, well positively correlated (0.67) with the highest scoring NL points also exhibiting high CP. Finally, Fig.~\ref{Fig:FurtherMetrics}g shows the CP as a function of MC, which has a pronounced negative correlation (-0.68).

\section{Additional ac susceptibility data for \coznmn{}} 
\label{SEC:coznmnPhase}

\begin{figure}[]
\centering
\includegraphics[width=0.5\linewidth]{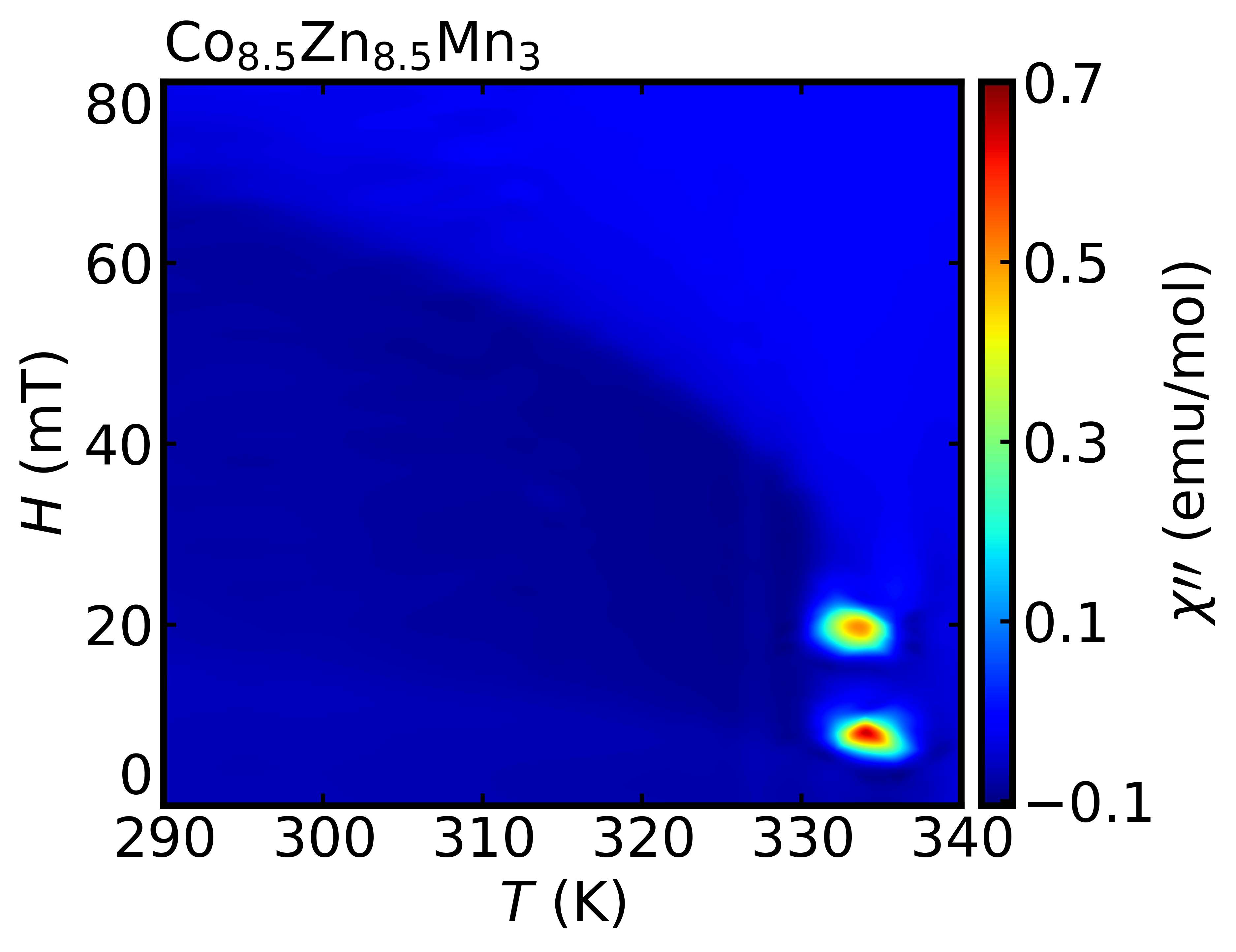}
\caption{\rev{A 2D plot of the imaginary-part of ac susceptibility ($\chi$'') of the Co$_{8.5}$Zn$_{8.5}$Mn$_{3}$ crystal.}}
\label{FIG:ImPhase_CoZnMn}
\end{figure}

\rev{Magnetic susceptibility measurements are sensitive to magnetic phase changes. We performed ac susceptibility experiments for \coznmn{}, yielding the real ($\chi '$; see Fig.~5a in main text) and imaginary ($\chi ''$) components. Figure~\ref{FIG:ImPhase_CoZnMn} shows $\chi ''$ of \coznmn{} with clear bright-regions in 330~<~$T$~<~340 and $H$~<~25, highlighting the presence of the skyrmion phase due to a slow relaxation process observed around the skyrmion phase\cite{Qian_PRB2016}.}

\section{Task-adaptable physical reservoir computing on FeGe}
\label{SEC:PRC_FeGe}

\begin{figure}
\centering
\includegraphics[width=1\linewidth]{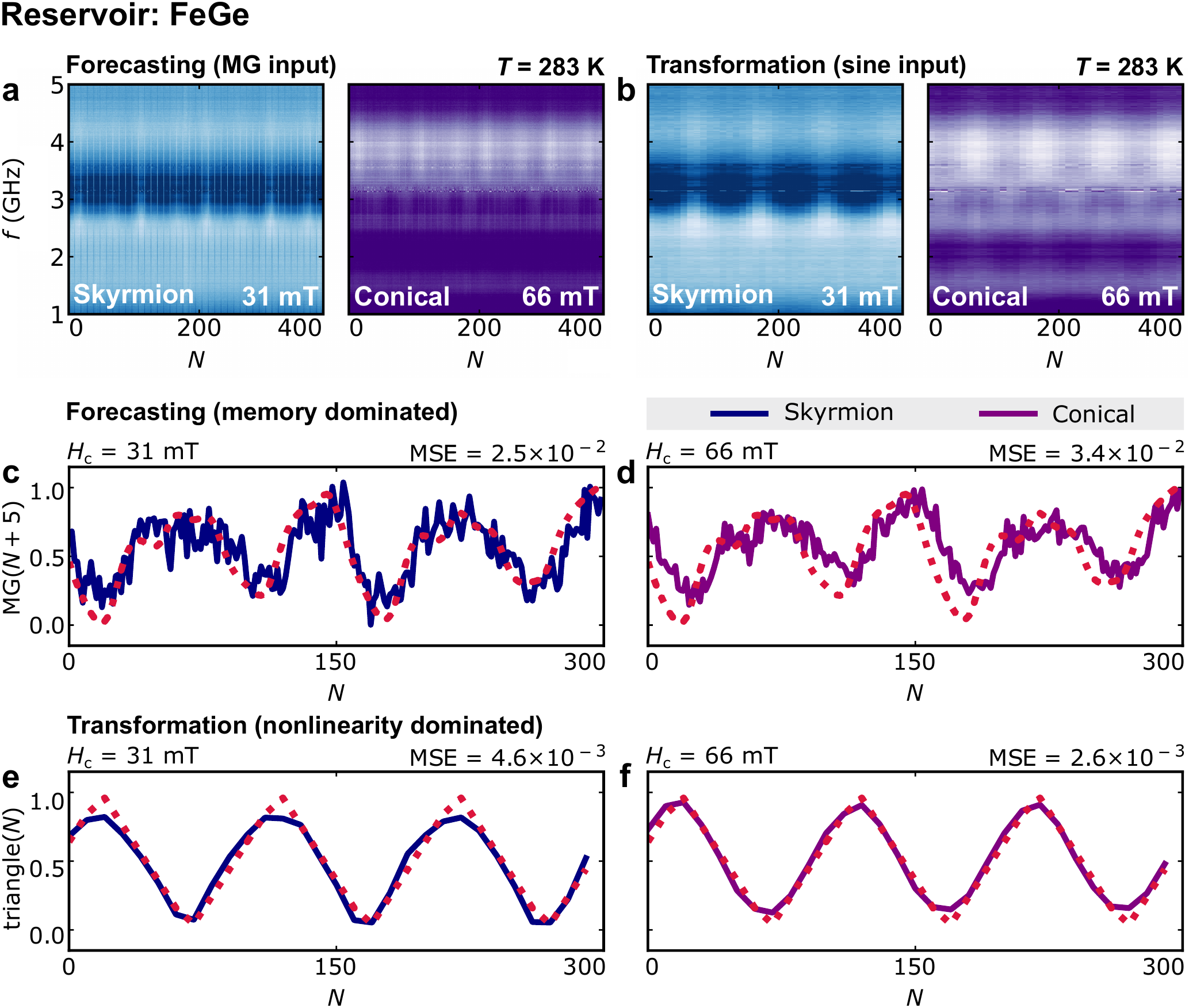}
\caption{\rev{Task-adaptive physical reservoir computing using FeGe at $T$~=~283~K. \textbf{a}, 2D plots of spin dynamics spectra measured as an evolution of $N$ for a Mackey-Glass and sine input functions at different values of $H_{\rm c}$ (31 and 66~mT). \textbf{b}\&\textbf{c}, Reservoir computing performance for predicting the MG function with 5 future steps and transforming a sine input signal to a triangle output, respectively. The dotted curves/lines represent the target function, while the solid curves/lines demonstrate the success of our task-adaptive physical reservoir computing approach.}}
\label{Fig:RC_FeGe}
\end{figure}

\rev{Similar to \cuoseo{} and \coznmn{} presented in this study, the chiral-lattice of FeGe also hosts a rich magnetic phase diagram, including skyrmions at near room temperature~\cite{yu_NatMat_2011,Porter_PRB_2014,Takagi_PRB_2017}. The following section summarises the phase-tunable approach using the FeGe sample at $T=283$~K with 5~mT cycling width. In Figs.~\ref{Fig:RC_FeGe}a\&b, we present the magnetic resonance spectra during field cycling for Mackey-Glass and sine input functions, which were used to perform future prediction of MG($N+5$) and transformation (sine to triangle) tasks shown in Figs.~\ref{Fig:RC_FeGe}c-f. Evidently, the spectra strongly depend on the choice of \Hc{}, highlighting the phase-tunability of physical reservoirs in this material system.} \rev{In particular, for the forecasting tasks (Figs.~\ref{Fig:RC_FeGe}c\&d), the skyrmion-dominated reservoir ($H_\text{c}$~=~31~mT) surpasses the conical reservoir ($H_\text{c}$~=~66~mT) in terms of MSE score (skyrmion: $2.5\times 10^{-2}$ vs conical: $3.4\times 10^{-2}$). However, for the transformation task, the conical reservoir achieves a better MSE than the skyrmion-dominated reservoir (skyrmion: $4.6\times 10^{-3}$ vs $2.6\times 10^{-3}$). Providing further evidence of phase-tunability of achieving task-adaptability.   
}

\section{High-temperature task-adaptability}
\label{SEC:HT_PRC}

\begin{figure}
\centering
\includegraphics[width=1\linewidth]{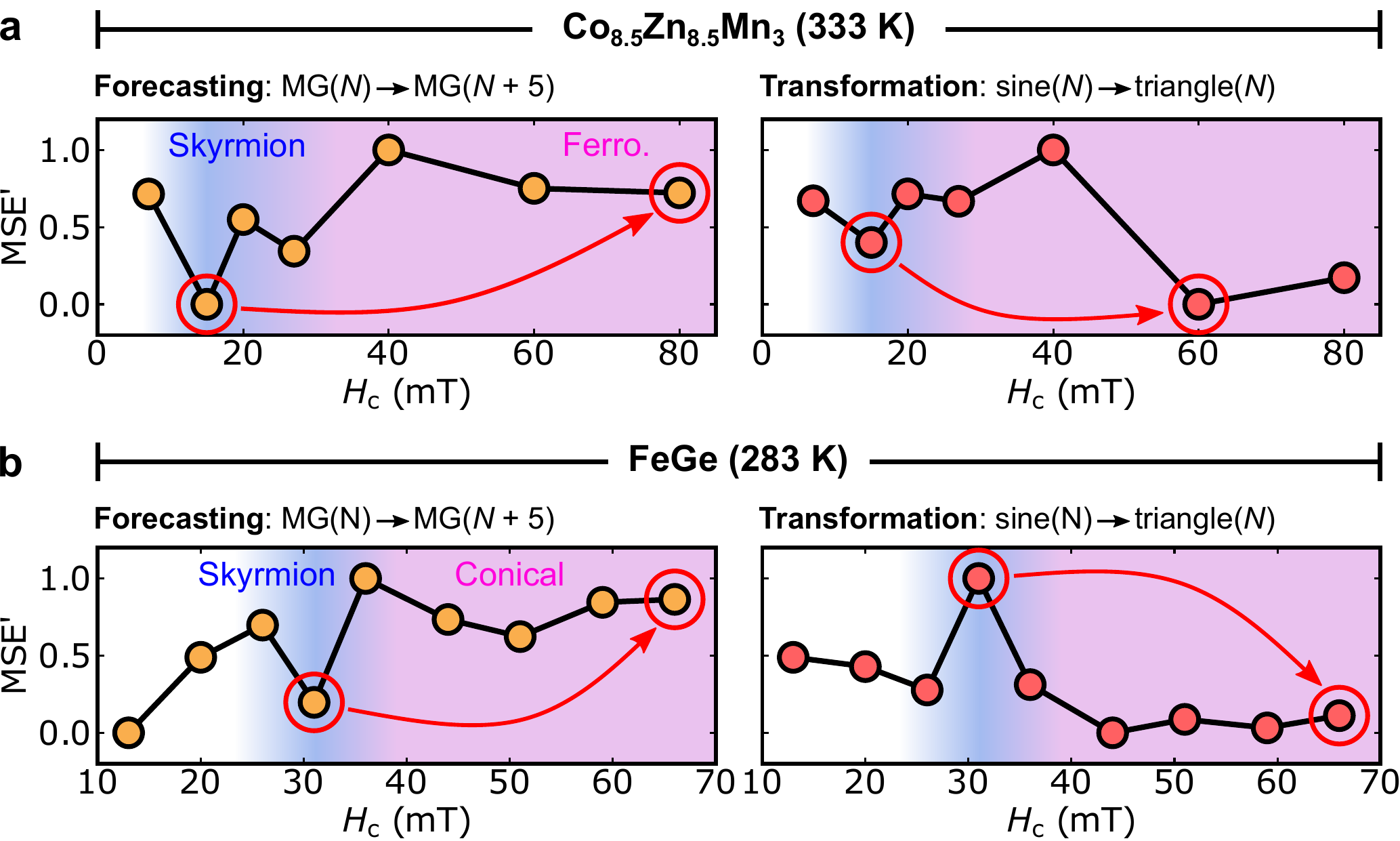}
\caption{\rev{\textbf{a}\&\textbf{b}, Comparison of task-adaptive physical reservoir computing using different materials and temperatures. Performance of forecasting a MG signal of five future steps and transforming a sine input signal to a triangle output function for Co$_{8.5}$Zn$_{8.5}$Mn$_{3}$ at $T$~=~333~K (\textbf{a}) and FeGe at $T$~=~283~K (\textbf{b}) as an evolution of $H_{\rm c}$. Blue and purple backgrounds denote the skyrmion and ferromagnetic/conical regions.}}
\label{Fig:RT_RC}
\end{figure}

\rev{In Fig.~\ref{Fig:RT_RC}, we demonstrate the phase-tunability of chiral magnets including \coznmn{} and FeGe near room temperature for two distinct tasks: forecasting and transformation. Similarly to Fig.~4b in the main text for \cuoseo{} at 4~K, the task prediction performance is plotted against \Hc{}. For forecasting tasks with \coznmn{} at 333~K (Fig.~\ref{Fig:RT_RC}a), the skyrmion phase exhibits the best performance at \Hc{}~=~15~mT, which gradually decreases as the system transitions through the conical phase and into the ferromagnetic state. Conversely, for transformation tasks, the performance improves when moving from the skyrmion phase to the ferromagnetic phase (e.g., \Hc{}~=~15 to 60~mT), highlighting the ability of the system to transform a sine input function into a triangular wave output. The same behaviour maintains persistence for the FeGe sample at 283~K (Fig.~\ref{Fig:RT_RC}b), i.e., the forecasting is best at the skyrmion phase (\Hc{}~=~31~mT) and decrease with increasing \Hc{}, and vice versa for transformation tasks. These further support that the task-adaptive reservoir computing concept can be transferable to a wide range of different materials.}

\color{black}{}

\clearpage

\bibliography{sn-bibliography}

%% file: Setup_files/Definitions.tex
\newcommand{\coznmn}{Co$_{8.5}$Zn$_{8.5}$Mn$_{3}$}

\newcommand{\Hc}{$H_{\rm c}$}